\documentclass[10pt, aps,prc,twocolumn,superscriptaddress,preprintnumbers,
amsmath, 
floatfix,
longbibliography,
nofootinbib
]{revtex4-1}
\usepackage[T1]{fontenc}
\usepackage[utf8x]{inputenc} 
\usepackage{adjustbox}          % Used to adjust figure frames
\usepackage[caption=false]{subfig}
\usepackage{url}
\usepackage{color}
\usepackage{float}
\usepackage[pdftex,colorlinks=true, linkcolor = blue, citecolor=blue,urlcolor=blue, bookmarksnumbered=true, bookmarksopen=true]{hyperref}
\usepackage{longtable}
\usepackage{amsfonts}
\usepackage{dsfont}
\usepackage{wrapfig,bm} 
\usepackage[normalem]{ulem}
\usepackage{MnSymbol}
\usepackage{float}
\usepackage{rotating}

\usepackage{xcolor}

\newcommand{\beq}{\begin{equation}}
\newcommand{\eeq}{\end{equation}}
\newcommand{\bea}{\begin{eqnarray}}
\newcommand{\eea}{\end{eqnarray}}

\begin{document}
\title{Time-Dependent Density Functional Theory Description of $^{238}$U(n,f), $^{240,242}$Pu(n,f) and $^{237}$Np(n,f) Reactions}

  \author{Aurel Bulgac}% 
%\email{bulgac@uw.edu}%
\affiliation{Department of Physics,%
  University of Washington, Seattle, Washington 98195--1560, USA} 
\author{Ibrahim Abdurrahman}%
%\email{ia4021kmm@gmail.com}
\affiliation{Theoretical Division, Los Alamos National Laboratory, Los Alamos, NM 87545, USA} 
\author{Matthew Kafker}%
%\email{kafkem@uw.edu}
\affiliation{Department of Physics,%
  University of Washington, Seattle, Washington 98195--1560, USA}   
 \author{Ionel Stetcu}  
\affiliation{Theoretical Division, Los Alamos National Laboratory, Los Alamos, NM 87545, USA}
\date{\today}

\begin{abstract}

In nuclei with an odd nucleon number the non-vanishing spin number density  is the
source of a pseudo-magnetic field, which favors the splitting of the nucleon Cooper pairs. Such an
pseudo-magnetic field is generated always in the dynamics of any nucleus, but its effects on Cooper
pairs is significantly enhanced in the dynamic evolution of nuclei with an odd number of nucleons.   
We present for the first time a microscopic study of the induced fission of the odd neutron compound nuclei 
$^{239}$U, $^{241, 243}$Pu, and the odd proton, odd neutron compound nucleus $^{238}$Np, performed within 
the time-dependent density functional theory extended to superfluid fermion systems, without any simplifying 
assumptions, with controlled numerical  approximations, and for a very large number of initial 
conditions. Due to the presence of the unpaired odd nucleon(s), 
the time-reversal symmetry of the fission compound nucleus is 
spontaneously broken, an aspect routinely neglected in the most advanced microscopic approaches of the past.
The emerging fission fragment properties are quite similar to the properties of fission fragments of 
neighboring even-even nuclei. The time from saddle-to-scission is often significantly longer in odd-odd 
or odd-mass nuclei than for even-even nuclei, since systems with unpaired nucleons are easier to excite and the 
potential energy surfaces of these nuclei have more structure, often resembling a very complicated
obstacle course, rather than a more direct evolution of the nuclear shape from the top of the outer 
fission barrier to the scission configuration. The Pauli blocking approximation, often invoked in the 
literature, expected to inhibit the fission of nuclei with unpaired nucleons,  is surprisingly strongly 
violated during the fission dynamics. 

\end{abstract} 

\preprint{NT@UW-25-2, LA-UR-25-21369v2}

\maketitle  

 \vspace{0.5cm}

Nuclei with an odd number of neutrons or/and protons have much higher level densities at the same 
excitation energy than those with even numbers of both nucleon types. This is true not only near the 
ground state, but also along the often-times circuitous path a nucleus takes from a compact shape to 
fission.  This includes navigation through the critical barrier regions and onto ultimate 
scission.  The oddness of the individual nucleon numbers leads to closely spaced potential energy surfaces 
and thus to an enhanced opportunity to jump from one surface to another.  This is similar to the situation 
known for a long time from chemistry studies~\cite{Born:1927,Tully:1990,Bulgac:2020}.  
For the excitation energies of the compound nuclei in induced fission with low energy 
neutrons~\cite{Bethe:1936,Bohr:1969}, even near the top of the outer fission barrier, where the nucleus is 
``cold'' according to 1956 Aage Bohr's analysis~\cite{Vandenbosch:1973}, and the number of accessible fission 
channels is still large in odd-mass or odd-odd nuclei.
The theoretical difficulties in dealing with 
this problem has lead to a dearth of treatments of the fission of systems with oddness, with 
simplifications and assumptions not sufficiently well underpinned by theory. 
The importance of the oddness issue was manifest with the first reactor produced plutonium for the 
Manhattan Project. 
While the microscopic amounts of pure $^{239}$Pu were produced with a cyclotron, which
has negligible spontaneous fission, the reactor produced also $^{240}$Pu, which has sufficient spontaneous 
fission that pre-detonation became a concern, that required implosion to circumvent.  
This is not an isolated case, the systematics of spontaneous fission half-lives of 
even-even (E-E) versus odd-even (O-E), even-odd (E-O), and odd-odd (O-O) compound nuclei 
is presented in every
discussion of fission~\cite{Bohr:1939,Hill:1953,Vandenbosch:1973,Wagemans:1991}.  The spontaneous fission 
half-lives of even-even nuclei are several orders of magnitude shorter than the spontaneous fission half-
lives of E-O, O-E or O-O nuclei, even though the fission barriers are very similar~\cite{Vandenbosch:1973}.

In nuclei with odd numbers of neutrons or/and protons, new order parameter(s) appear(s) in the static mean 
fields, proportional to the spatial spin number density (a quantity similar to a local spin 
magnetization) and a pseudo-magnetic field emerges
${\bf B}({\bf r}) \propto {\bf s}({\bf r})$~\cite{Bender:2003}, 
which impacts the nucleon spins in a matter similar to a magnetic field acting on a spin 1/2-particle
${\bf B}({\bf r}) \cdot{\bm \sigma}\propto {\bf s}({\bf r}) \cdot{\bm \sigma}$.
These new pseudo-magnetic fields are axial vectors, which 
break both the rotational and time-reversal symmetry. Such pseudo-magnetic fields appear also in the
dynamic evolution of E-E systems, e.g. in induced fission, which
is a highly non-equilibrium/dissipative
process~\cite{Bulgac:2016,Bulgac:2019c,Bulgac:2020,Shi:2020}. As we demonstrate in this work, in the case of 
induced fission of compound nuclei with an odd number of neutrons or/and protons, 
their effect is significantly stronger than in the case of induced fission of E-E compound nuclei. 
As established by \textcite{Caldeira:1983}, tunneling times are dramatically 
affected by dissipation, and it is expected that dissipation will manifest itself differently 
in nuclei with even and odd nucleons. The time-reversal properties of the many-body wave functions are also 
crucial in describing $\beta$-decay of fission fragments (FFs) and $P$-even and $T$-odd asymmetries
of angular correlations of FFs
with light particles emitted in induced fission with cold polarized neutrons~\cite{Lyubashevsky:2025}. 
 
The microscopic treatment of the low-energy structure and dynamics of odd-$A$ and O-O nuclei has 
always presented significant technical difficulties compared to E-E nuclei.  
An extra odd nucleon cannot participate in a Cooper pair, and the corresponding single-particle state 
can not contribute to the anomalous density 
(the superfluid order parameter)~\cite{Bohr:1958,Bulgac:2024a,supplement}, 
thus inhibiting pairing correlations compared to neighboring even-even nuclei.
It has been established microscopically, in full agreement with the experimental findings, that the induced 
fission dynamics of E-E nuclei has a strongly non-equilibrium 
character~\cite{Bulgac:2016,Bulgac:2019c,Bulgac:2020}, now accepted by the community \cite{Bender:2020}. At 
the same time, pairing correlations play a crucial 
role in the evolution of the nuclear shape of the compound nucleus~\cite{Bertsch:1980,Barranco:1988,Barranco:1990,Bertsch:1991,Bertsch:1994,Bertsch:1997}. 
In the absence of pairing correlations, fission dynamics is strongly slowed and often brought to a stop, 
while by artificially increasing their strength, the dynamics is accelerated 
by at least an order of magnitude and the dynamics is mostly adiabatic~\cite{Bulgac:2019c}.

In the literature, two different approximations are typically used to treat pairing in nuclei with odd 
nucleon numbers is static calculations,
the Pauli blocking approximation~\cite{Ring:2004} 
and the equal filling approximation (EFA)~\cite{Perez-Martin:2008}.  
In the presence of pairing correlations, the many-body wave functions of systems with even or odd numbers of
protons/neutrons do not have a well-defined particle number in the
Hartree-Fock-Bogoliubov (HFB) mean-field approximation. 
In the Pauli blocking approximation the corresponding quasiparticle level of the odd nucleon does not 
contribute to the anomalous density, but it does contribute to all 
the other one-body number densities, and this leads to the 
spontaneous breaking of the time-reversal and particle number symmetries, and along with them also the 
rotational symmetry is also often broken. On the other hand, in the EFA the nucleus is treated as an even-
even nucleus, with exactly vanishing pseudo-magnetic fields  
${\bf B}({\bf r})\propto {\bf s}({\bf r}) \equiv 0$ 
and no time-symmetry breaking terms,
but with an average particle number corresponding to the corresponding odd number of 
nucleons~\cite{Perez-Martin:2008}. 
%There is a clear difference between the Pauli blocking approximation  and the EFA. 
In the EFA pair correlations are expected to be stronger than in the Pauli blocking approximation, since the odd nucleon(s) contribute to the pairing correlations in the EFA.

\begin{figure}
\includegraphics[width=1.0\columnwidth]{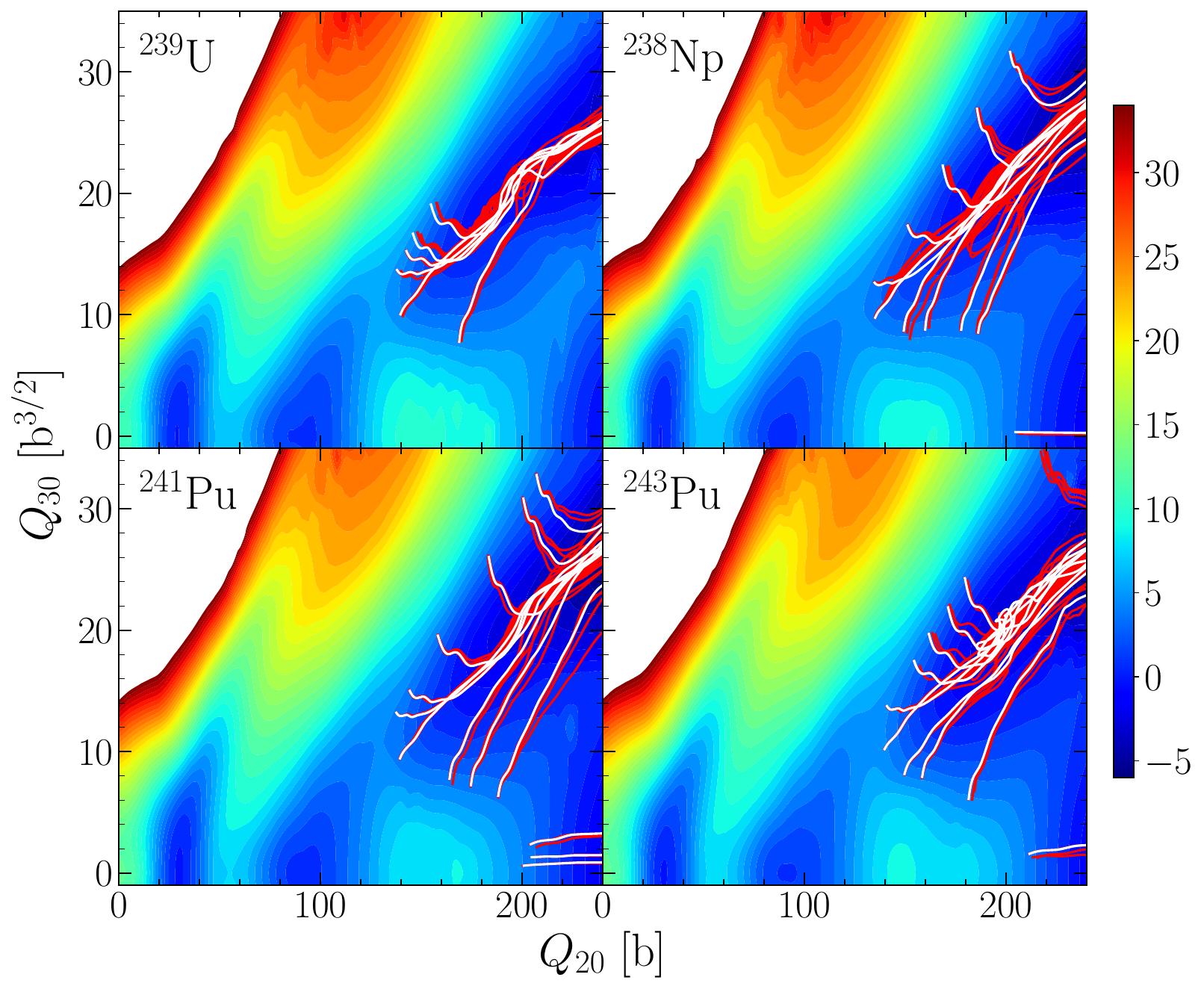}
\caption{\label{fig:zero}   For the four nuclei with odd number of protons and/or neutrons reported here we show 
the individual TDSLDA trajectories for the even-even seed nuclei (white) and for the odd-nucleon nuclei (red) 
on the initial PES of the ``seed'' even-even nucleus. The symmetric trajectories, for relatively small initial $Q_{30}(t=0)$
extend beyond the limits of each plot and they are showed in Fig.~\ref{fig:two}. The initial conditions have been chosen along the rim of the outer fission barrier as in previous studies \cite{Bulgac:2016,Bulgac:2019c,Bulgac:2020,Bulgac:2020a,Bulgac:2021,Scamps:2023a}. }
\end{figure}

\begin{figure}
\includegraphics[width=0.8\columnwidth]{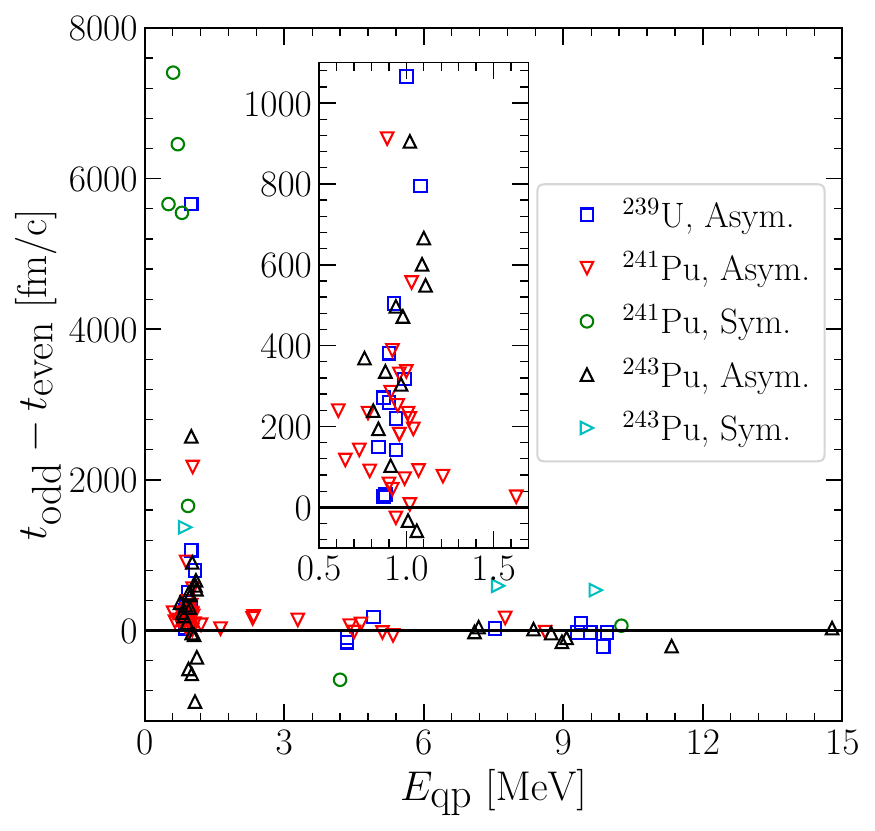}
\caption{ Differences in evolution times between the odd-nucleon systems 
and the corresponding even-even ``seed'' nuclei. States with initial blocked nucleon(s) with $E_{qp} < 1$ MeV
often show distinctively longer evolution times, in agreement with the trajectories illustrated in Fig.~\ref{fig:two}. 
See also the detailed Tables in SOM \cite{supplement}, which also includes a similar figure for $^{237}$Np. }\label{fig:000} 
\end{figure}

 \begin{table*}[!htb]
\centering
\caption{  \label{tb:table}  The TKE, the FF proton and mass numbers, 
and the number of runs perform in each case and their variance in parentheses. 
We categorize the runs as ``symmetric'' and ``asymmetric'' depending on the initial value 
of the octupole deformation of the compound nucleus $Q_{30}$ being smaller 
or larger than $\approx 2.5$ b$^{3/2}$. Next to each nucleus symbol O$-$O, O$-$E, E$-$E stand  for 
the odd and respectively even particle numbers of neutrons and protons, and 
the subscript S or A stands for symmetric and asymmetric. In case of $^{235}$U(n,f) and 
$^{239}$Pu(n,f) we have added the corresponding TKE and  the FF proton and mass numbers 
obtained in our previous papers~\cite{Bulgac:2021,Bulgac:2022b,Scamps:2023a}, but never reported until now. Our results for TKE are within 1 MeV from Madland systematics for $^{235,238}$U(n,f) and $^{239}$Pu(n,f) \cite{Madland:2006}. Average properties of FFs can be compared against \textsc{cgmf} results \cite{supplement}, which in turn are based on available experimental data~\cite{TALOU2021108087}. A direct comparison with experimental data for all observables is challenging given the corrections that enter into the experimental analysis to extract the pre-neutron emission information.} %See also Ref. \cite{supplement} for additional information.}
%\begin{tabular}{p{2.5cm} p{2.5cm} p{2.5cm} p{2.5cm} p{2.5cm} p{2.5cm} p{1.8cm}}
\begin{tabular}{ccccccccc}
\hline \hline
& Treatment & Type & TKE & $Z_H$ & $Z_L$ & $A_H$ & $A_L$  & No. Runs \\
\hline
$^{235}$U(n,f) & E$-$E & asym. & 168.71(4.03) & 52.07(0.84) & 39.94(0.84) & 134.85(2.24) & 101.16(2.24) & 22 \\ 
$^{235}$U(n,f) & E$-$E & sym. & 142.35(3.05) & 47.36(1.35) & 44.64(1.35) & 121.60(3.58) & 114.40(3.58) &  2\\ 
\hline
$^{237}$Np(n,f) & O$-$O & asym. & 174.68(4.34) & 52.34(0.56) & 40.66(0.53) & 135.11(1.45) & 102.90(1.35) & 46\\ 
$^{237}$Np(n,f) & E$-$E & asym. & 175.36(2.75) & 52.18(0.27) & 40.84(0.27) & 134.52(0.73) & 103.50(0.72) & 10 \\ 
$^{237}$Np(n,f) & O$-$O  & sym. & 151.04(1.30) & 49.53(0.05) & 43.50(0.05) & 127.37(0.05) & 110.69(0.02) & 2\\ 
$^{237}$Np(n,f) & E$-$E & sym. & 148.24           & 49.90           & 43.10          & 128.36          & 109.66           & 1 \\ 
\hline
$^{238}$U(n,f)& O$-$E & asym. & 169.08(3.56) & 51.85(0.49) & 40.15(0.49) & 135.43(1.29) & 103.54(1.32) & 23 \\ 
$^{238}$U(n,f)& E$-$E & asym. & 169.40(1.36) & 51.60(0.32) & 40.40(0.32) & 134.79(0.80) & 104.21(0.80) & 6 \\ 
\hline
$^{239}$Pu(n,f)& E$-$E & asym. & 176.85(1.00) & 53.09(0.21) & 40.91(0.21) & 136.49(0.43) & 103.51(0.43) & 5 \\ 
$^{239}$Pu(n,f)& E$-$E & sym. & 150.03(1.16) & 49.87(1.02) & 44.13(1.02) & 128.46(2.57) & 111.54(2.57) & 2  \\ 
\hline
$^{240}$Pu(n,f)& O$-$E & asym. & 173.88(4.18) & 52.71(0.75) & 41.29(0.75) & 136.03(1.79) & 104.93(1.80) & 36 \\ 
$^{240}$Pu(n,f)& E$-$E & asym. & 173.98(3.00) & 52.79(0.75) & 41.21(0.75) & 136.15(1.66) & 104.85(1.66) & 10 \\ 
$^{240}$Pu(n,f)& O$-$E & sym. & 151.58(3.59) & 50.70(0.76) & 43.19(0.75) & 131.11(1.92) & 109.69(1.76) &  7\\ 
$^{240}$Pu(n,f)& E$-$E & sym. & 151.26(2.64) & 51.00(0.42) & 42.88(0.47) & 131.48(1.23) & 109.41(1.27) & 3 \\ 
\hline
$^{242}$Pu(n,f)& O$-$E & asym. & 171.50(4.96) & 52.51(0.61) & 41.49(0.61) & 136.62(1.41) & 106.40(1.42) & 32 \\ 
$^{242}$Pu(n,f)& E$-$E & asym. & 170.04(2.99) & 52.58(0.59) & 41.42(0.60) & 136.78(1.40) & 106.21(1.42) & 9 \\ 
$^{242}$Pu(n,f)& O$-$E & sym.  & 151.41(1.38) & 50.78(0.34) & 43.22(0.34) & 132.54(0.83) & 110.48(1.00) & 3 \\
$^{242}$Pu(n,f)& E$-$E & sym. & 151.62        & 50.72       & 43.28       & 131.95     
             & 111.05       & 1 \\
\hline \hline
\end{tabular}

\label{table:initial}
\end{table*}

The most recent and detailed microscopic descriptions of the fission of odd-$A$ nuclei were obtained within the EFA  
and were reported in Refs.~\cite{Schunck:2023, Pore:2024}, where brief reviews of older and simplified theoretical studies can be found. 
In Refs.~\cite{Schunck:2023, Pore:2024} odd-nucleon systems have been treated within the density functional theory (DFT) framework as even-even within the EFA approximation, thus with unbroken time-reversal symmetry and even particle parity. One might argue that the inaccuracy of adopting either the Pauli blocking approximation or the EFA is of the order 
$1/N$ or $1/Z$  respectively, but this is in clear conflict with at least one observable, in particular the fact that
spontaneous fission half-lives of even-even on odd-nucleon nuclei differ by orders of magnitude~\cite{Vandenbosch:1973}. 
The single-particle occupation probabilities are double degenerate in odd-nucleon systems in the static calculations, with the exception of the Pauli blocked state. Under time evolution, these doublets are no longer double degenerate, due to the presence of the induced pseudo-magnetic fields, see the supplemental online material (SOM)~\cite{supplement}. The effect of these induced pseudo-magnetic fields is also significantly more pronounced in compound nuclei with odd number of nucleons than in E-E compound systems.

\begin{figure*} 
\includegraphics[width=0.8\columnwidth]{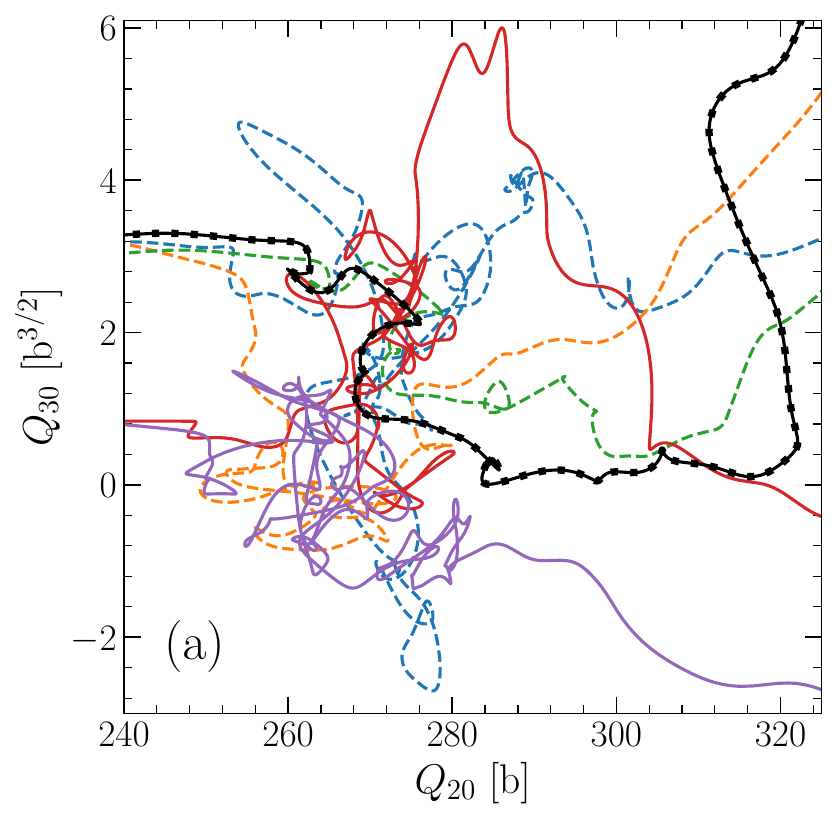} 
\includegraphics[width=0.8\columnwidth]{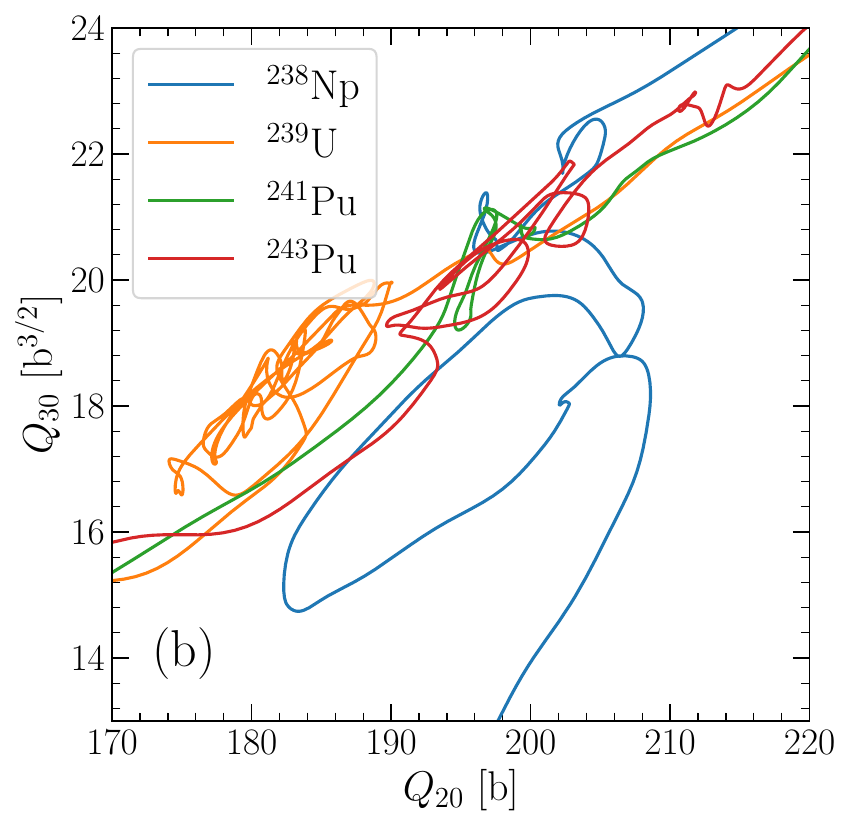} 
\caption{ \label{fig:two} Selected TDSLDA trajectories with long saddle-to-scission time from Fig. 1. 
Panel (a) shows data for $^{241}$Pu only. These trajectories have
low initial octupole deformations $Q_{30}$ and, as a rule, low final TKE of the FFs. 
In panel (a) the trajectory for one ``seed'' E-E nucleus $^{241}$Pu is shown with a black-with-dots line.  Panel (b) 
shows trajectories where the initial $Q_{30}$ is not small. In all cases where the trajectory in $Q_{20}$ and 
$Q_{30}$ is ``convoluted,'' the saddle-to-scission time is significantly longer than for ``smoother'' 
trajectories with similar initial deformations. The time from scission until full FFs separation 
is always very short ${\cal O}(100)$ fm/c.}
\end{figure*}

\begin{figure}
\includegraphics[width=0.9\columnwidth]{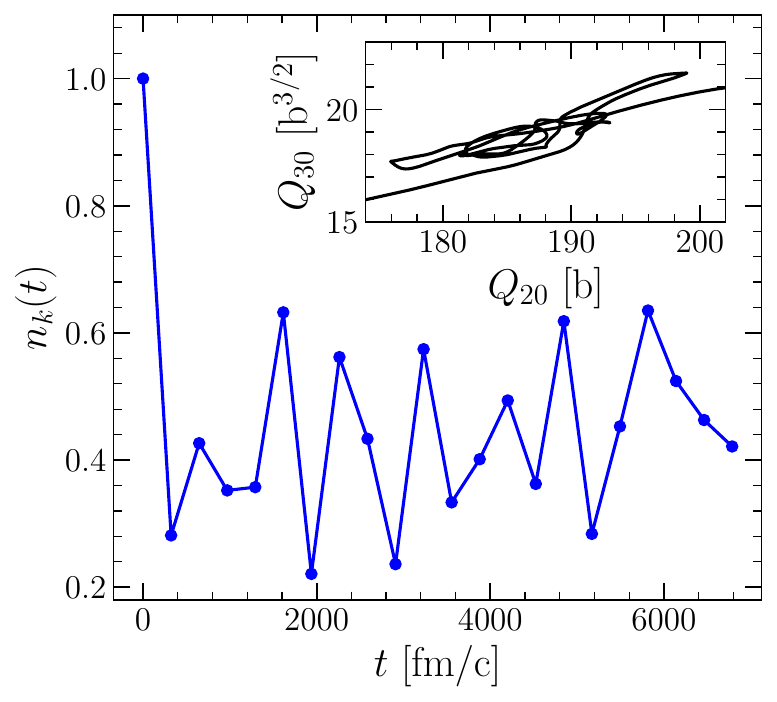}
\caption{ \label{fig:three}  The occupation probability $n_k(t) =\langle v_k(t)|v_k(t)\rangle$ for a chosen 
Pauli blocked quasiparticle state in $^{239}$U. In this simulation we have used canonical quasiparticle wave functions initially, see Refs.~\cite{supplement,Bulgac:2024a}.  Under time evolution however the quasiparticle wave functions cease to be canonical. The inset displays the shape evolution of the compound nucleus in the time interval 280-5000 fm/c.  }
\end{figure}

\begin{figure}
\includegraphics[width=0.9\columnwidth]{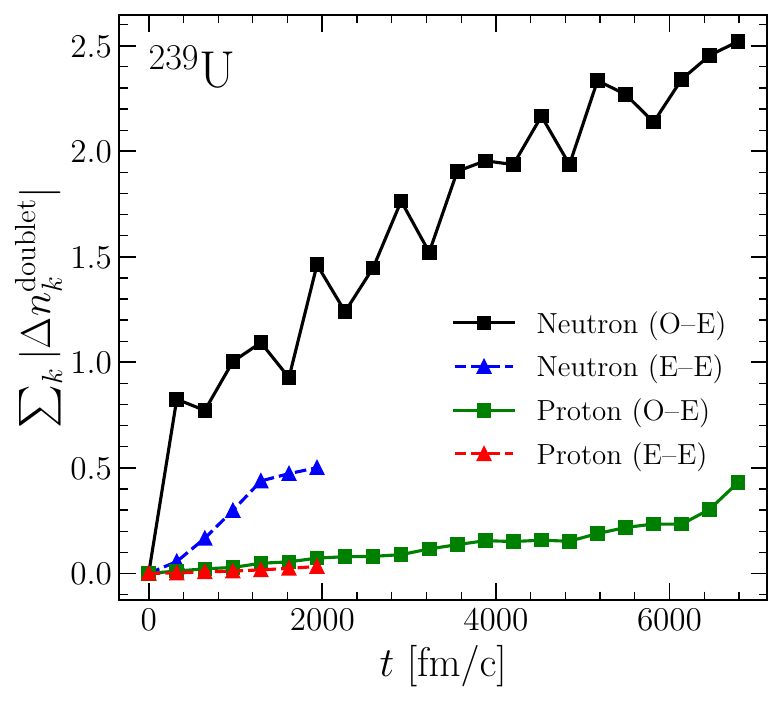}
\caption{ \label{fig:sumdoublets} The sum of the absolute difference of the nucleon occupation probabilities 
between ``Kramers doublets'' as a function of time for the system treated as O-E/O-O and 
E-E, due to the role of time-odd fields.   The run starts in 
the canonical basis and the neutron level with $n_k(0)\equiv 1$ is excluded.}
\end{figure} 

In this work we present the induced fission of odd-$A$ compound nuclei 
$^{239}$U (23 trajectories), $^{241}$Pu (43 trajectories), $^{243}$Pu (35 trajectories) 
and of odd-odd nucleus $^{238}$Np (48 trajectories) within time-dependent density functional theory extended to include 
pairing correlations, dubbed the time-dependent superfluid local density approximation (TDSLDA)~\cite{Bulgac:2013a,Bulgac:2019},
without any assumptions or uncontrolled approximations. 
In total we have simulated 149 trajectories for these nuclei, considering various initial conditions corresponding 
to using different quasiparticle states for the odd nucleon and various initial quadrupole and octupole deformations $(Q_{20}, Q_{30})$
of the initial nucleus outside the outer fission barrier~\cite{supplement}. The initial states were generated by considering a ``seed'' even-even (E-E) nucleus with the average $N$ and $Z$ identical to the corresponding to the target odd mass (O-E) or odd-odd (O-O) nucleus, and thus the volume, surface, Coulomb, and symmetry energies are essentially identical. The only difference between these nuclei left being the small change(s) in the pairing gap(s) and the particle parity of the corresponding many-body wave functions~\cite{supplement}.
In order to contrast the time evolution of even-even 
 nuclei versus those with an odd number of protons and/or neutrons, for each set of initial deformations $Q_{20},\, Q_{30}$ of the  ``seed'' nuclei we analyzed the 
corresponding fission dynamics of even-even ``seed'' nuclei $^{239}$U (6 trajectories), $^{241}$Pu (13 trajectories), $^{243}$Pu (10 trajectories), 
and $^{238}$Np (11 trajectories), see Fig.~\ref{fig:zero} and SOM~\cite{supplement}. 
We used then the prescription outlined in 
Refs.~\cite{Ring:2004,Bertsch:2009,Schunck:2023,Pore:2024} and flipped various  low energy quasiparticle 
states with occupation probability $n_k\approx 0.5\pm0.2$ to create odd-nucleon systems, with a small number of exceptions; see SOM~\cite{supplement}, which contains also an additional number of Refs~\cite{Bardeen:1957,Bulgac:1980,Tan:2008b,Zwerger:2011,Dobaczewski:1984,Belyaev:1984,Dobaczewski:1996,Abdurrahman:2024,Shi:2020,Bulgac:2018,Marevic:2022,Scamps:2012,Ren:2022,Ren:2022a,Tong:2022,Bulgac:2024,Feynman:1955}.  

The energy density functional (EDF) SeaLL1~\cite{Bulgac:2018} used here is a typical modern 
accurate EDF, constructed to describe overwhelmingly even-even nuclei only, 
see Ref.~\cite{Zurek:2024} for one of the most recent studies and 
for earlier references. Most of the time-reversal symmetry
contributions to an EDF however arise from the presence of an odd number of nucleons and their 
magnitudes are known only from a limited number of studies~\cite{Bender:2003,Grams:2023,Bulgac:2018}.
Thus, one can expect that the actual role of time-reversal symmetry-breaking terms in induced fission in
particular could be significantly larger than what we find here.

When comparing the FF properties for $^{238}$U(n,f), $^{240, \, 242}$Pu(n,f), and $^{237}$Np(n,f) reactions to 
those emerging from fission of E-E compound nuclei, remarkable differences can be observed in Table \ref{table:initial}.
 O-E and O-O systems have total kinetic energy (TKE) distributions about twice as broad as those of E-E nuclei
$^{236}$U, $^{240}$Pu or ``seed'' E-E nuclei $^{239}$U, $^{241, \, 243}$Pu or $^{238}$Np. 
Similar differences can also be observed for the average neutron and proton number distributions in the fission fragments.  Thus, a correct treatment of the time-reversal symmetry breaking is essential to obtain the correct distribution width.

When considering 
$^{239}$U, $^{241,\,243}$Pu, and $^{238}$Np nuclei we have chosen a wide range of quasiparticle states in the ``seed'' nuclei,
resulting in a wide distribution of the initial energy of the corresponding odd-nucleon nuclei. The largest
variations in the TKE is between initial states with either rather small or large initial octupole deformations $Q_{30}(0)$. These choices of 
$Q_{30}(0)$  lead to predominantly symmetric-like FF distributions in the case of small initial $Q_{30}(0)$ with smaller TKE, 
similar to what we have observed in the induced fission of even-even nuclei~\cite{Bulgac:2019c}. In particular, the trajectories originating from 
initial states with low $Q_{30}(0)$ often appear to be ``confused,'' as the system might wander around both positive 
and negative values of $Q_{30}$. If the initial value of $Q_{30}(0)$ 
($Q_{30}(t)\approx \langle \Phi(t)| z(5z^2-3r^2)|\Phi(t) \rangle >0.0$) 
relatively large and positive, 
the heavy fission fragment typically ends up on the left ($Q_{30}(t_\textrm{final})>0$)  
on the fission $Oz$-axis in the intrinsic reference frame 
of the fissioning compound nucleus.
For low initial values of $Q_{30}(0)$ close to zero 
the system might take some time, before deciding that the HFF should go to the left or right (namely to 
$z<0$ or $z>0$, $z$ being the fission axis), and that usually leads to longer saddle-to-scission 
evolution times, particularly in the case of relatively low quasiparticle energies, 
see Figs.~\ref{fig:000}, \ref{fig:two} and the SOM~\cite{supplement}. 
In case of low-energy quasiparticle states chosen for the odd system, 
the fissioning nucleus has a relatively low intrinsic initial energy, and that also corresponds to a 
significantly large total level density of the nucleus with one or two odd nucleons when compared to an even-
even system 
at comparable energies at and above the outer fission barrier. This leads to a significantly larger 
number of potential energy surfaces (PESs) per unit energy, and to an increased role of dissipative effects
manifested by more frequent jumps between these more 
closely spaced PESs as the trajectory departs from the initial state. 
This behavior has been observed in Ref.~\cite{Bulgac:2019d}, where the TDSLDA has been augmented to include 
the otherwise absent stochastic elements. 
Otherwise, the evolution times from  saddle-to-scission are
very similar for both E-E and odd-nucleon compound nuclei, see Fig.~\ref{fig:000} and Fig. 3 in the 
SOM~\cite{supplement}.

A salient aspect of our results is illustrated in Fig.~\ref{fig:three}, demonstrating that the Pauli 
blocking approximation is strongly violated. Due to the emergence of pseudo-magnetic effective fields,  
a nucleus with odd nucleons 
experiences repopulations of the quasiparticle states and the trajectory 
is quite long, similar to those illustrated in Fig.~\ref{fig:two}.   
We have constructed the canonical states as described in Refs.~\cite{Bulgac:2023,Bulgac:2024a} 
many-body wave functions at all times. Only one quasiparticle state at any time $t\neq 0$
has unit occupancy and is nondegenerate in the case of one odd neutron in $^{239}$U, $^{242, 243}$Pu and 
for both nucleon types in $^{238}$Np. The quasiparticle occupation probabilities emerging 
from static DFT calculations are different from the canonical occupation probabilities, 
and only the latter display the expected behavior of a Pauli blocked state, see  Refs.~\cite{Bulgac:2023,supplement,Bulgac:2024a} for details. Even if one chooses initial
canonical wave functions, their canonical character is lost for all times $t>0.$
Only in static DFT/HFB calculations the generalized density 
matrix commutes with the generalized mean field and
then for each quasiparticle state one can assign a quasiparticle energy $E_k$ 
and a corresponding occupation probability $n_k = \langle v_k|v_k\rangle$. We illustrate in
Fig.~\ref{fig:sumdoublets} the significantly enhanced role of the dynamically generated 
pseudo-magnetic field(s), which break(s) time-reversal symmetry during the fission dynamics of nuclei 
with an odd-nucleon number, as compared to the case of an E-E nucleus, which is the real cause the 
Pauli blocking approximation is violated and why also the EFA is also expected to be inaccurate.

\vspace{0.5cm}
We thank L. G. Sobotka for critical reading of the manuscript and valuable suggestions.
A.B. was supported by the Office of Science, Grant No. DE-FG02-97ER41014  
and partially by NNSA cooperative Agreement DE-NA0004150. 
M.K. was supported by NNSA cooperative Agreement DE-NA0003841.
I.A. and I.S. were supported by the U.S.
Department of Energy through the Los Alamos National
Laboratory. The Los Alamos National Laboratory is operated
by Triad National Security, LLC, for the National Nuclear
Security Administration of the U.S. Department of
Energy Contract No. 89233218CNA000001.
This research used resources of the Oak Ridge Leadership Computing 
Facility, which is a U.S. DOE Office of
Science User Facility supported under Contract No. DE-AC05-00OR22725. I.A. and I.S. gratefully acknowledge partial support and computational
resources provided by the Advanced Simulation and
Computing (ASC) Program. \\

Data availability - The data are not publicly available.
The data are available from the authors upon reasonable
request.  
 
  %%
%% These are needed to avoid a babel error.
\providecommand{\selectlanguage}[1]{}
\renewcommand{\selectlanguage}[1]{}

\bibliography{local_fission}  
  
\end{document}

\title{Time-Dependent Density Functional Theory Description of $^{238}$U(n,f), $^{240,242}$Pu(n,f) and $^{237}$Np(n,f) Reactions\\
Online Supplemental Material}

  \author{Aurel Bulgac}% 
%\email{bulgac@uw.edu}%
\affiliation{Department of Physics,%
  University of Washington, Seattle, Washington 98195--1560, USA} 
\author{Ibrahim Abdurrahman}%
%\email{ia4021kmm@gmail.com}
\affiliation{Theoretical Division, Los Alamos National Laboratory, Los Alamos, NM 87545, USA} 
\author{Matthew Kafker}%
%\email{kafkem@uw.edu}
\affiliation{Department of Physics,%
  University of Washington, Seattle, Washington 98195--1560, USA}   
 \author{Ionel Stetcu}  
\affiliation{Theoretical Division, Los Alamos National Laboratory, Los Alamos, NM 87545, USA}
\date{\today}

\begin{abstract}
This document contains supplementary material for the paper "Time-Dependent Density Functional Theory Description of $^{238}$U(n,f), $^{240,242}$Pu(n,f) and $^{237}$Np(n,f) Reactions".
\end{abstract} 

\preprint{NT@UW-25-2, LA-UR-25-21369}

\maketitle  

 \vspace{0.5cm}

 \section{Structure of static and time-dependent mean field equations for even-even versus odd nuclei} \label{ap:A}
 
 In the static mean field description of pairing there are several aspects which distinguishes the 
treatment of even-even, odd-even, even-odd, and odd-odd nuclei. The many-body wave function of an even-even 
nucleus is invariant with respect to two distinct transformations, one in real space and the other 
in the gauge space. In both cases the wave function remains unchanged under a rotation by an angle of $2\pi$ . On the other hand, 
the many-body wave function under a rotation by the same angle $2\pi$ in both the real and the 
gauge space the many-body wave function for systems with an odd number of fermions acquires the phase $\exp(i\pi)=-1$. An odd-odd 
nucleus obviously does not change sign under any of these rotations.  The many-body wave functions of even-even systems of stationary states remain unchanged under time-reversal symmetry, 
which is not true anymore in systems with an odd number of fermions.
A correct description of $\beta$-decays can be achieved only of 
time-reversal symmetry is properly incorporated in the many-body wave functions.  
Time-reversal symmetry and parity are responsible for the $P$-even and $T$-odd angular correlations of light particles with FFs emitted in induced fission with cold polarized neutrons~\cite{Lyubashevsky:2025}. 

A static mean field many-body wave function for an even-even nucleus is constructed 
in two different ways, which are not equivalent. The most general approach is using quasiparticle 
single-particle annihilation operators $\alpha_k$. This approach was considered recently in Refs.~\cite{Bulgac:2023,Bulgac:2024a}, 
where a number of  aspects not described previously in the literature of such a wave function are described as well.
\begin{align}
|\Phi\rangle = {\cal N} \prod_k\alpha_k|0\rangle, \label{eq:hfb}
\end{align}
where $|0\rangle$ is the true  vacuum, ${\cal N}$ a normalization factor,
\begin{align}
& \alpha^\dagger_k =\sumint_\xi[ u_k(\xi)\psi^\dagger (\xi) + v_k (\xi) \psi(\xi)], \label{eq:HFB-1}\\
&\alpha_k = \sumint_\xi [v^*_k(\xi) \psi^\dagger(\xi) + u^*_k(\xi)\psi(\xi)], \\
&\sumint_\xi |u_k(\xi)|^2 +|v_k(\xi)|^2 =1,
\end{align}     
 where $\xi$ stands for spatial/spin/isospin coordinates of a nucleon and $\psi^\dagger(\xi),\,\psi(\xi)$ 
 are field creation and annihilation operators.

 A related form of  the many-body wave functions 
 is the Bardeen-Cooper-Schrieffer (BCS)  form~\cite{Bardeen:1957}
 \begin{align}
 &|\Phi\rangle = \prod_k (u_k + v_k a^\dagger_k  a^\dagger_{\overline {k}})|0\rangle,\quad {\rm where} \quad  u_k^2+v_k^2=1, \label{eq:bcs}\\
 &\langle 0 | a_ka^\dagger_k|0\rangle = \langle 0| a_{\overline{k}} a^\dagger_{\overline {k}} | 0 \rangle =1,
 \quad \phi_{k,{\overline{k}}} (\xi ) =\langle \xi | a^\dagger_{k,\overline{k}}|0\rangle 
  \end{align}
where $a^\dagger_k \, a^\dagger_{\overline k}$ are creation operators for time-reversed single-particle states.
This type of wave function is sufficiently accurate when the pairing gap is relatively small ($\Delta \ll \varepsilon_F$) and 
 the corresponding strength of the pairing correlations are weak, a situation one typically encounters in metals.
For nuclear systems and cold atomic gases this is not the case anymore and the 
appropriate choice of a many-body wave function is Eq.~\eqref{eq:hfb}. Unfortunately, the BCS type of many-body wave 
functions is widely used in nuclear calculations, 
with the justification that the correction due to pairing correlations 
to the total energy is relatively small. If that would be a correct argument 
then a many-electron wave functions would be good enough without pairing correlations, and 
the relative correction to the binding energy would be much smaller than in the case of nuclear systems, 
but the structure would be completely wrong, as a superconductor is in a highly correlated phase.
As shown in Ref.~\cite{Bulgac:1980}, in the case of a local pairing field, 
as typical in Kohn-Sham implementations of DFT in nuclear physics, the anomalous density always has a divergence of the type
\begin{align}
\kappa(\xi,\xi') =\langle\Phi|\psi(\xi')\psi(\xi)|\Phi\rangle \propto 
\frac{1}{| {\bm r}_1-{\bm r}_2|},
\end{align}
when $| {\bm r}_1-{\bm r}_2| \rightarrow 0$,  
which leads to a power-like behavior of the fermion distributions~\cite{Tan:2008b,Zwerger:2011}
\begin{align}
n(p) \propto \frac{C}{p^4} \quad {\rm when} \quad p\rightarrow \infty,
\end{align}
a behavior which cannot be reproduced with a BCS many-body wave function, see discussion 
in Refs.~\cite{Bulgac:2023,Bulgac:2024a} in the case of nuclear systems.

The quasiparticle single-particle wave functions $[u_k(\xi),\,v_k(\xi)]$ are eigenvectors of the static SLDA/HFB Hamiltonian, 
while the single-particle wave functions $\phi_{k,{\overline{k}}} (\xi )$ are usually the eigenstates
of an (approximate) mean field Hamiltonian.
There is no question that the many-body wave functions defined by Eq.~\eqref{eq:hfb} are better 
approximations than the wave functions defined by Eq.~\eqref{eq:bcs}, especially in the case of nuclear systems.

If one were 
to simulate a nucleus in a box of size $L^2\times2L\approx 2\times 30^3$ fm with a lattice constant $l=L/N=1$ fm, the number of either proton or neutron quasiparticle states in Eq.~\eqref{eq:hfb} is given by,
\begin{align}
& N_{box}= 2\times60\times30^2=108,000, \\ 
&k_{cut}=\frac{\pi}{l},
\end{align}
for either the proton or neutron subsystem, while for the choice of the many-body wave function in Eq.~\eqref{eq:bcs} the corresponding number is 
\begin{align}
& N_{nucleus} \approx 2  \frac{4\pi}{3}r_0^2A \frac{4\pi}{3} k_0^3 \times\frac{1}{(2\pi)^3}\approx \frac{16A}{9\pi}\approx 0.56 A, \label{eq:BCS}\\ 
&\quad k_o =\sqrt{\frac{2 m |V_0|}{\hbar^2}},
\end{align} 
and $V_0\approx -50 $ MeV is the depth of the single-particle mean field~\cite{Bohr:1969}. 
The choice of a lattice constant $l=1$ fm corresponds to a nucleon momentum cutoff $p_{cutt}\approx \hbar \pi/l\approx 600$ MeV/c, comparable to the $\Lambda_{QCD}$ and thus suitable to handle even chiral effective field theory nucleon interactions.
The reason why $N_{nucleus} \ll N_{box}$ is pretty simple, the nuclear BCS many-body wave function is constructed only of 
bound nucleon levels, which unfortunately for single-particle energies $\varepsilon_k >\mu$ have 
also wrong asymptotic tails~\cite{Bulgac:1980,Dobaczewski:1984,Belyaev:1984,Dobaczewski:1996}, which 
simultaneously affects the spatial behavior of the pairing gap potential. 
When the nucleus is approaching either the proton or neutron drip line and the 
Fermi energy is equal to $|V_0|$ there are no available bound states  to form a Cooper pair. 

\begin{figure}[h]
\includegraphics[width=0.85\columnwidth]{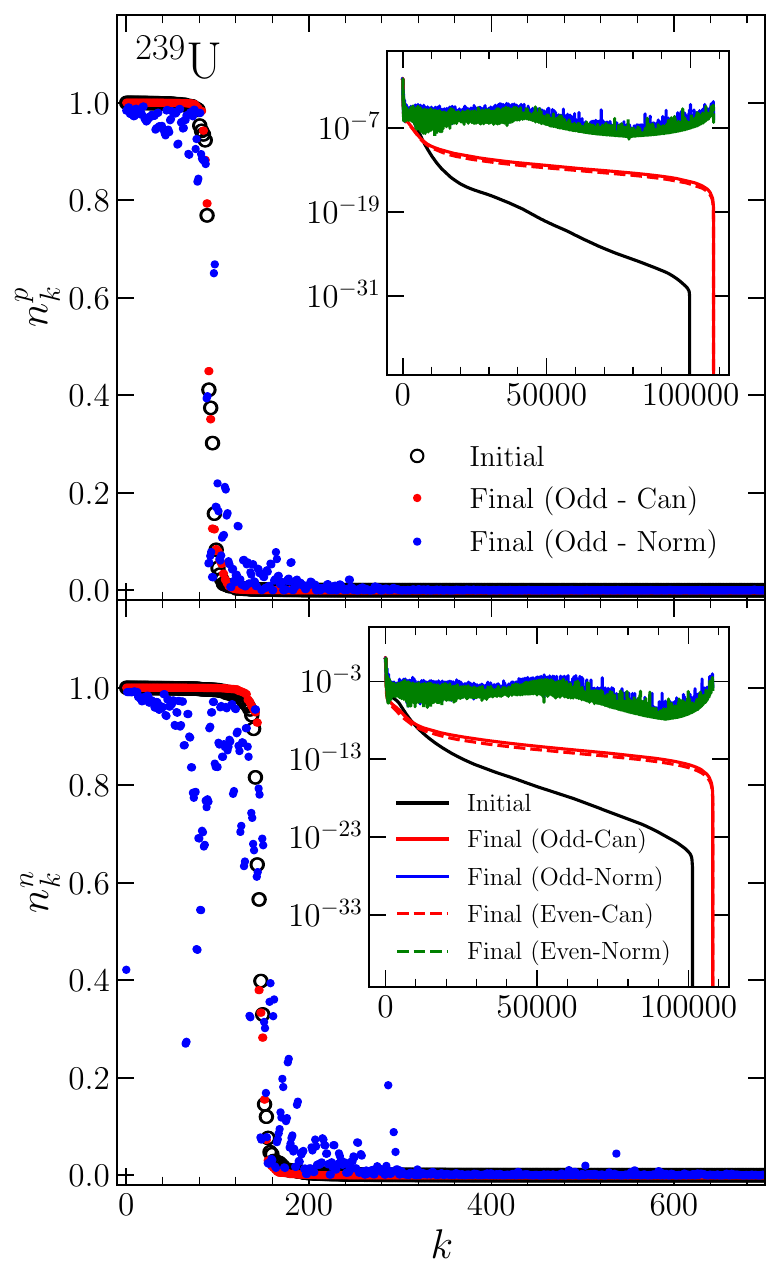}
\caption{ \label{fig:occs} The initial/final occupation number distributions in the canonical representation as well as the final occupation number distribution in the normal representation for protons (top panel) and neutrons (bottom panel) are shown. In the inset the same quantities are shown in a logarithmic scale for the initial and final states of the simulation. The legend in the top panel refers to both of the linear occupation distributions, while the legend in the bottom inset refers to both logarithmic occupation probabilities (contained in the insets).  
} 
\end{figure}

In Refs.~\cite{Bulgac:2023,Bulgac:2024a} one can find a rather detailed study of the nature of time-reversed orbitals participating in a Cooper pair.
In reality the number of orbitals participating in the formation of Cooper pairs in nuclei is much larger than the upper limit given by Eq.\eqref{eq:BCS}. 
However, most studies performed in the literature always assume a BCS form of the many-body wave function 
for a nucleus with pairing correlations. In Ref.~\cite{Bulgac:2024a} we have presented arguments 
for why using a BCS many-body wave function can lead to large errors, 
particularly in the case of time-dependent phenomena, such as induced  fission. TDSLDA simulations can be
performed using for the initial state either quasiparticle single-particle wave functions obtained from a
static SLDA/HFB calculation or canonical single-particle wave functions with undistinguishable results. 
However, during the time evolution, due to the presence of naturally arising time-odd fields at subsequent 
times the quasiparticle and canonical occupation probabilities are not identical, even if the initial set of 
single-particle states was canonical at $t=0$ fm/c, compare at the final time 
the normal/quasiparticle and canonical occupation probabilities in Fig.~\ref{fig:occs} and also the similar 
results presented in Refs.~\cite{Bulgac:2023a,Bulgac:2024a}.

In the case of an odd number of nucleons the many-body wave function for either the proton or neutron subsystem 
is often chosen as~\cite{Ring:2004,Bertsch:2009,Schunck:2023,Pore:2024}
\begin{align}
|\Phi\rangle = {\cal N}\alpha_l^\dagger\prod_k\alpha_k|0\rangle, \label{eq:hfb-1}
\end{align}
or correspondingly as
 \begin{align}
& |\Phi\rangle = a^\dagger_l\prod_{k\neq l} (u_k + v_k a^\dagger_k  a^\dagger_{\overline {k}})|0\rangle, \label{eq:bcs-o1}\\
& |\Phi\rangle = a^\dagger_{\overline{l}}\prod_{k\neq l}  (u_k + v_k a^\dagger_k  a^\dagger_{\overline {k}})|0\rangle. \label{eq:bcs-o2}
\end{align}

In order to construct systems with odd number of nucleons we first chose a ``seed'' even-even nucleus with the same $N$ and $Z$ as the target compound nucleus.
For example, if we want to simulate the induced fission of compound nucleus $^{239}$U, we construct a shape constraint state for an even-even  
fictitious nucleus with $Z=92$ and $N=147$ nucleons.  We typically choose, with a few exceptions, in this ``seed'' nucleus a quasiparticle state with occupation probability
$n_k=0.5\pm0.2$ \textcolor{blue}, as in that case the atomic number of the intended odd-mass nucleus is $N=147+(1-2n_k)\approx147\pm0.4$ \textcolor{blue}, see detailed tables in the Appendix. 
For the ``seed'' even-even nucleus we also perform induced fission simulations, 
specifically 6 runs for compound systems $^{239}$U, 13 runs for $^{241}$Pu, 2 runs for $^{243}$Pu, and 11 runs for $^{238}$Np, corresponding to the selected set of 
initial deformations  $(Q_{20}, Q_{30})$, chosen along the rim of the outer fission barrier. 
We also added, for comparison, 
unpublished results for the induced fission of $^{236}$U (24 runs) and $^{240}$Pu (7 runs), for which we 
reported only 
results for the FF spin properties in the past~\cite{Bulgac:2021,Bulgac:2022b,Scamps:2023a}.
The implementation of the rule Eq.~\eqref{eq:hfb-1} is realized by swapping one chosen quasiparticle wave function for a single 
chosen index $k$ according to the rule~\cite{Ring:2004,Bertsch:2009}
\begin{align}
(u_k(\xi), v_k(\xi)) \leftrightarrow (v^*_k(\xi), u^*_k(\xi)), \label{eq:ring}
\end{align}
in all one-body densities. In prior studies~\cite{Perez-Martin:2008,Schunck:2023,Pore:2024} this procedure was implemented, however, with the 
equal-filling-approximation, 
which artificially enforces time-reversal symmetry on the many-body wave function of the odd-nucleon system.
These types of many-body wave functions for odd systems have by definition an odd particle number parity, unlike the 
wave functions defined in Eqs.~(\ref{eq:hfb}, \ref{eq:bcs}) and acquire a phase $\exp(i\pi)=-1$ 
after a 3D spatial or gauge rotation around any axis.
A Pauli blocked quasiparticle single-particle state $\langle\xi | \alpha^\dagger_k |0\rangle$, with an initial canonical occupation probability $n_k(0) =\langle v_k(0)|v_k(0)\rangle\equiv 1$
evolves in time with the same quasiparticle Hamiltonian as the unblocked
quasiparticle states, but that does imply that the corresponding $u$-component 
is non-vanishing at later times, and thus for all $t>0$ 
$n_k(t) =\langle v_k(t)|v_k(t)\rangle< 1$,
see Fig.~(5) in the main text. 
This behavior is by default violated 
in any time-dependent Hartree-Fock (TDHF) + time-dependent BCS (TDBCS) approach for odd systems, as the 
Pauli blocked state will remain blocked at all times, which is another deficiency of using a TDBCS approach for odd systems, in addition to the
points discussed above. 
%\textcolor{orange}{(MK: At some point, shouldn't we mention that explicit antisymmetrization of the anomalous density is necessary to obtain good energy conservation, as IA demonstrated?)}

\begin{figure}
\includegraphics[width=0.9\columnwidth]{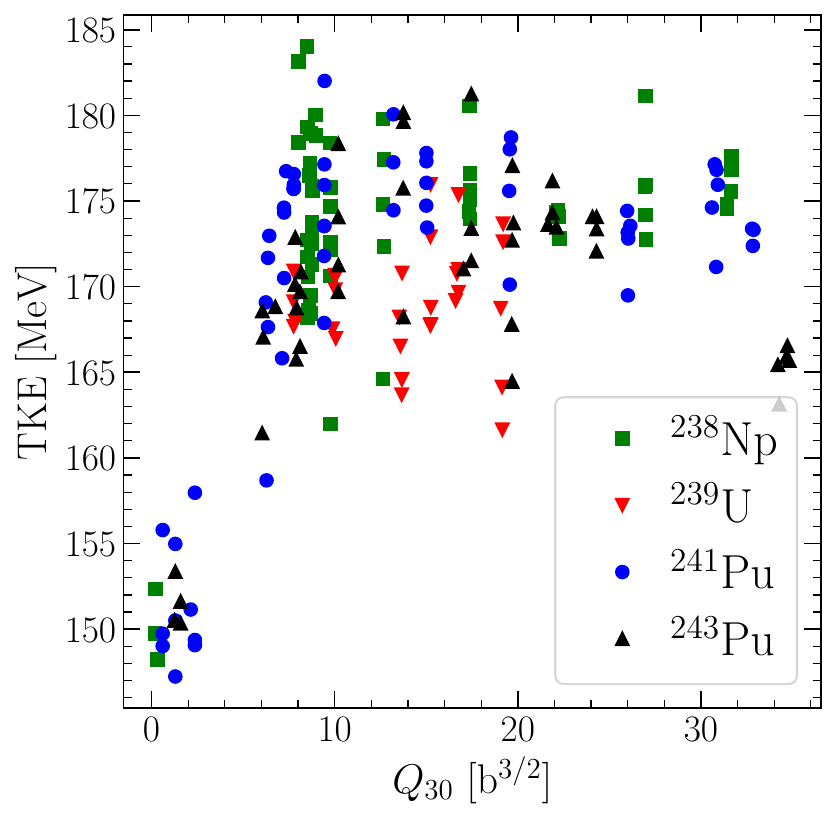}
\caption{ \label{fig:0000}   The final TKE as a function of the initial value of $Q_{30}$. 
Notice the cluster of initial states for $^{241}$Pu with TKE smaller  than 160 MeV discussed 
in the main text and illustrated as a function of time in Fig. 3 in the main text.}
\end{figure}

% \begin{figure}[h]
% \includegraphics[width=0.85\columnwidth]{sumndoublets.pdf}
% \caption{ \label{fig:sumdoublets} The sum of the absolute difference of the neutron occupation probabilities between  "doublets" as a function of time for both odd-nucleon and even-even systems.   All runs start in the canonical basis, and are recorded in both the normal and canonical basis in times after. For odd-fermion systems in this sum of ``doublets'' the one single-particle with canonical 
% occupation probability $n_k(t)\equiv 1$ at all time is excluded.
% Since in a odd-fermion system the mean field brakes the time-reversal symmetry at $t\equiv 0$ 
% and the many-body level density is higher in an odd-fermion system than in even-even system at low energies, the effect of time-symmetry breaking terms are significantly larger in odd-fermion systems.
% } 
% \end{figure} 

 The time-dependent simulations were performed in the same manner as in our previous studies of induced fission of even-even 
nuclei~\cite{Bulgac:2016,Bulgac:2019c,Bulgac:2020,Bulgac:2021,Bulgac:2022b,Scamps:2023a,Abdurrahman:2024,Shi:2021}, 
with the only difference being in the preparation of the initial state. 
The codes are available online and they are periodically updated on GitHub. We have generated fully self-consistent ground states within SLDA with the energy 
density functional SeaLL1~\cite{Bulgac:2018} for the above nuclei with the corresponding average $N$ and $Z$, 
treated however as even-even systems. We placed the nucleus in a simulation box $30^2\times60$ fm$^3$ 
with periodic boundary conditions and a lattice constant $l=1$ fm. The total number of the quasiparticle states is 
$2\times 2\times 30^2\times60 = 216,000$. We initialize the compound nucleus in neighborhood of the outer fission barrier and the corresponding 
self-consistent solution is found with specified quadruple and octupole deformations $Q_{20,}, \, Q_{30}$ first using the code HFBTHO~\cite{Marevic:2022}, followed by a  converged self-consistently solution on the lattice.
In order to generate the initial quasiparticle wave function for the odd-nucleon nuclei 
we have ``flipped'' various quasiparticle wave functions with occupation probabilities $n_k\approx 0.5\pm 0.2$, and relatively low quasiparticle 
energies $E_k < 10$ MeV, see Appendix B.
This procedure  ensures that the proton and/or neutron particle parities of the corresponding many-body wave functions are -1. 

In static SLDA/HFB calculations the generalized density matrix commutes with the generalized mean field and then for each quasiparticle state 
one can uniquely assign a quasiparticle energy $E_k$ and a corresponding occupation probability 
$n_k = \langle v_k|v_k\rangle$. 
In the case of TDSLDA the quasiparticle wave functions are no longer eigenstates of 
the generalized single-particle Hamiltonian.
An arbitrary unitary transformation of the quasiparticle wave functions $u_k(\xi),v_k(\xi)$, 
ocorrespondingly a unitary transformation
$\beta^\dagger_l = \sumint_k {\cal U}_{lk}\alpha^\dagger_k$, does not affect the normal one-body 
and the anomalous densities and it does not affect the many-body generalized Slater determinant.

In the typical textbook or BCS representation of the generalized Slater determinant, see Eq.~\eqref{eq:bcs}, 
the canonial occupation probabilities $n_k(t)=v_k^2(t),  u_k^2(t)+ v_k^2(t)=1$ for any $t>0$ 
are not identical to the occupation probabilities $\langle v_k(t)|v_k(t)\rangle$,  obtained using 
quasiparticle wave functions, see Eq.~\eqref{eq:HFB-1}.  The corresponding single particle wave functions used in defining many-body 
wave functions using canonical wave functions  $\phi_{k,\overline{k}}(\xi)$, see 
Eqs.~(\ref{eq:bcs}, \ref{eq:hfb-1}-\ref{eq:bcs-o2}), 
and the corresponding canonical occupation probabilities $n_k(t)=v_k^2(t)$
are the true canonical single-particle wave functions/natural orbitals~\cite{Bulgac:2023,Bulgac:2024a}, 
which are eigenstates of the single particle density matrix 
\begin{align}
& n(\xi,\xi',t) = \sumint_kv^*_k(\xi,t)v_k(\xi',t), \\
&\sumint_{\xi'} n(\xi,\xi',t)\phi_k(\xi',t)=v_k^2(t)\phi_k(\xi,t). \label{eq:can}
\end{align} 
The canonical single-particle wave functions provide the most economical representation of 
a generalized Slater determinant as a sum with the smallest number of 
$N$-particle Slater determinants~\cite{Bulgac:2023}. All one-body 
number and anomalous densities can be represented either 
through quasiparticle wave functions  or
through canonical wave functions and corresponding canonical 
occupation probabilities.

% \begin{figure}
% \includegraphics[width=0.85\columnwidth]{npair.pdf}
% \includegraphics[width=0.85\columnwidth]{CanOccsDiffSumVsTime.pdf}
% \caption{ \label{fig:six} The differences between the occupation probabilities of ``doublets'' in the canonical basis as a function of time for $^{239}$U treated as an even-nucleus (blue dots) and in the inset the time-evolution of the corresponding differences.  
%  In the lower panel we show the sum of 
% the absolute differences between the occupation probabilities of ``doublets'' in the canonical basis as a function of time for $^{239}$U. 
% } 
% \end{figure}

\begin{figure}
\includegraphics[width=0.9\columnwidth]{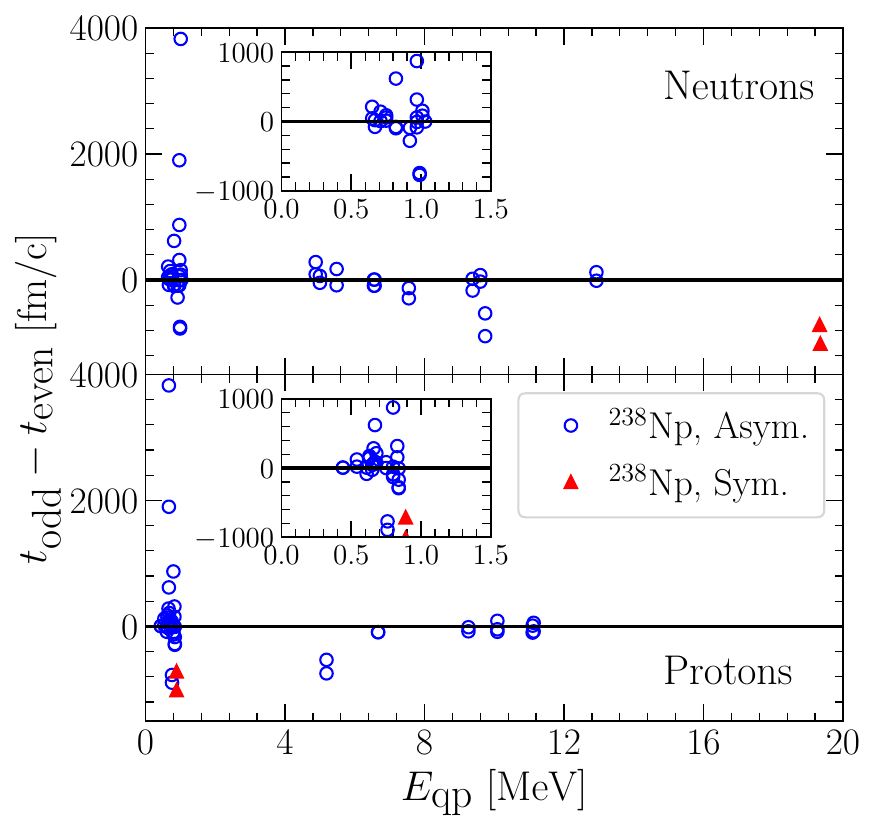}
\caption{ \label{fig:000}  The difference in evolution times between the odd-nucleon systems 
and the corresponding even-even ``seed'' nuclei. States with initial blocked nucleon(s) with $E_{qp} < 1$ MeV for $^{238}$Np. This figure is similar to Fig. 2 in the main paper.}
\end{figure}

While quasiparticle wave functions are evolved in time in TDSLDA simulations continuously, 
the canonical single-particle wave functions have to be determined 
by diagonalizing the time-dependent 
number density matrix $n(\xi,\xi',t)$ at each time step. 
Unlike in the TDBCS approximation a canonical single-particle state at 
one specific time $t$ ceases to be a canonical single-particle 
state when propagated in time, which is one of the uncontrollable 
assumptions and approximation used in TDDFT/TDHF+TDBCS
simulations. 
In other words TDDFT/TDHF+TDBCS 
is not mathematically equivalent to the mathematically well defined TDDFT 
extended to superfluid correlations or to TDHFB and the accuracy 
of the published so far results using TDDFT/TDHF+TDBCS is questionable. 
It is also well known that TDDFT/TDHF+TDBCS framework 
violates a fundamental equation of quantum fluid dynamics, 
the continuity equation~\cite{Scamps:2012}
\begin{align}
    \frac{d n(\xi,t)}{dt} + {\bm \nabla}\cdot {\bm j}(\xi,t) = 
    \sum_k \frac{d n_k(t)}{dt}|\phi_k(\xi,t)|^2\not \equiv 0, \label{eq:cont}
\end{align}
where $n(\xi,t)$ is the number density, ${\bf j}(\xi,t)$ is the current density, and $n_k(t)$ and $\phi_k(\xi,t)$ are the time-dependent occupation probabilities single-particle wave-functions used in TDDFT/TDHF+TDBCS simulations.
This equation has to be contrasted with the exact continuity equation
\begin{align}
    \frac{d n(\xi,t)}{dt} + {\bm \nabla}\cdot {\bm j}(\xi,t) = 0
\end{align}
which is valid both at classical and quantum many-body levels, and also in TDSLDA or TDHF or TDHFB 
frameworks. Refs.~\cite{Ren:2022,Ren:2022a,Tong:2022} are a few recent examples, where the incorrect 
continuity equation~\eqref{eq:cont} has been incorporated 
(apparently unknowingly in most cases) in the TDDFT fission simulations, which were
initiated by various authors in 2015. As demonstrated in Ref.~\cite{Bulgac:2024a}, 
see Section VI, the use of a treatment of paring correlations in basically any reduced Fock 
space leads to significant errors and basically irreproducible results by any approximation 
deemed reasonable otherwise. 

\begin{figure}  \includegraphics[width=0.9\columnwidth]{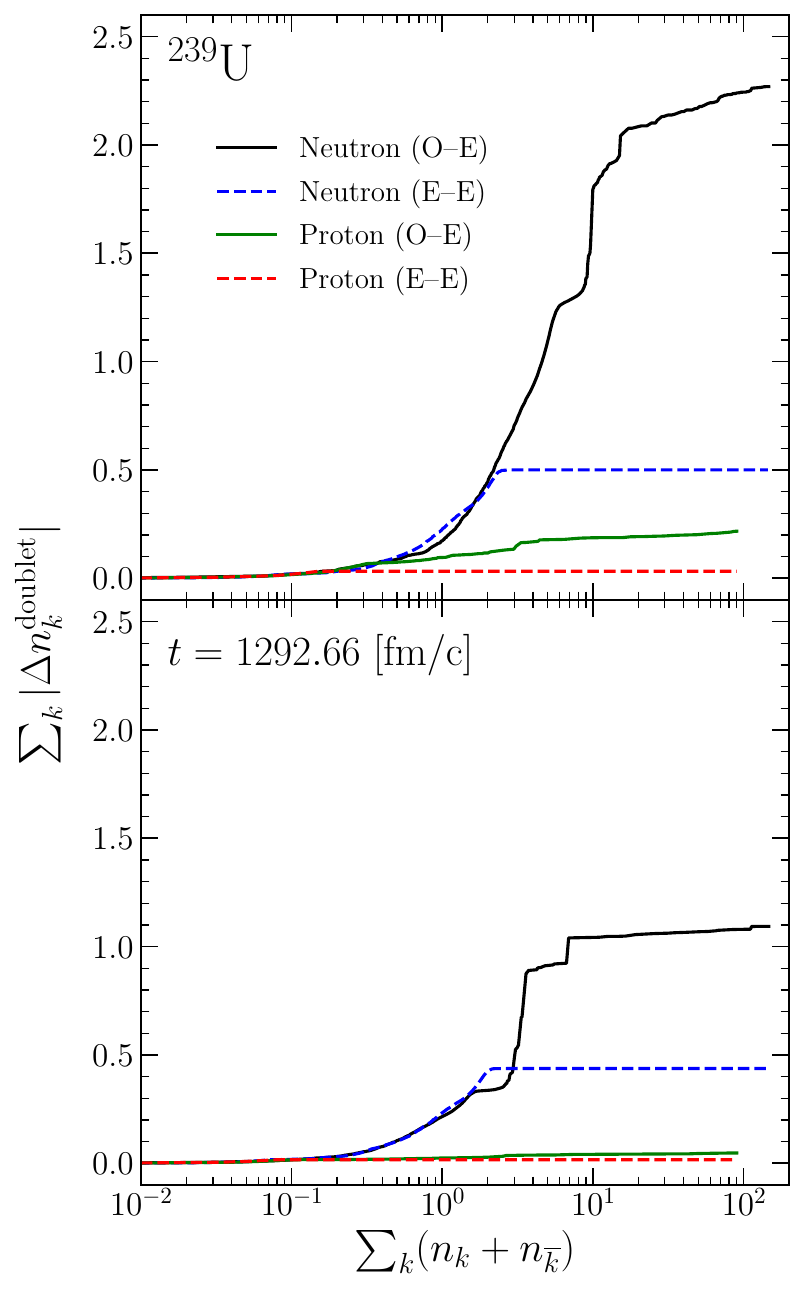} \caption{ \label{fig:trbreaking} The sum of the absolute difference of the nucleon occupation probabilities 
between ``doublets'' as a function of the cumulative sum of paired occupation numbers for the system treated as (O-E) and (E-E) starting from the lowest occupied pair.   The run starts in 
the canonical basis and the neutron level with $n_k(0)\equiv 1$ is excluded. The top panel shows the final time while the bottom panel shows an intermediate time before scission.  There is a marked difference between the dynamical evolution of an E-E system and a system with an odd number of neutrons in this case treated as an O-E systems as expected, when basically all doublet splittings are significantly affected by time symmetry breaking terms.}
\end{figure}

\begin{figure}  \includegraphics[width=0.9\columnwidth]{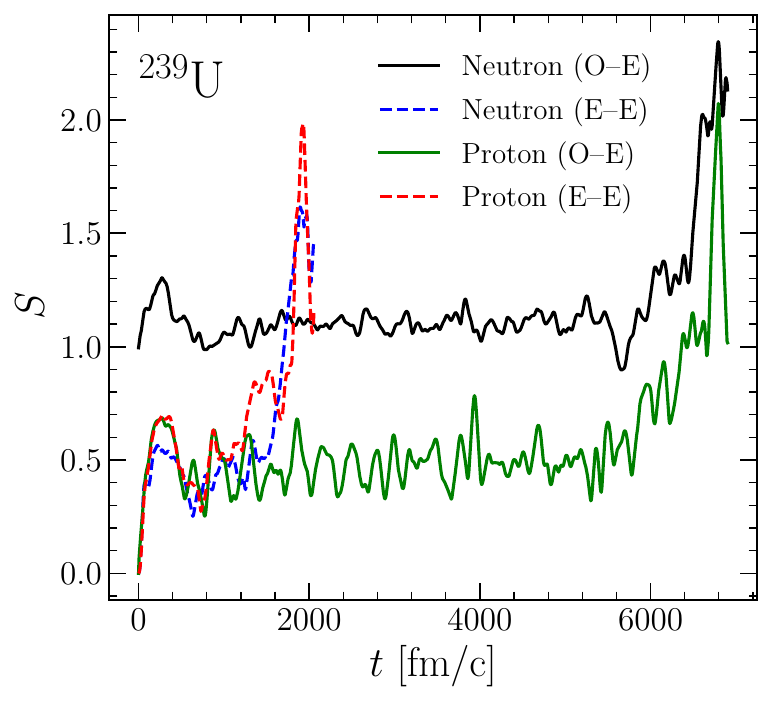} \caption{ \label{fig:sdens} The quantity $S = \int \sqrt{\bm{s}(\bm{r}) \cdot \bm{s} (\bm{r})} d^3 r $ is shown as a function of time for the system treated as (O-E) and (E-E). While the fact that for the O-E case fluctuates above 1 until scission is expected, the fact that the proton spin fluctuation fluctuates around the value $\approx 0.5$, similar to what one observes in the case of both neutrons and protons for a E-E system is a measure of how the time symmetry breaking effects are important in a dynamics evolution. Since after scission the FFs can have either an odd or even number of either protons or neutrons explains why the corresponding $S$ increases considerably in all cases.  }  
\end{figure}

In the Pauli blocking approximation the canonical occupation probability 
of the singled-out quasiparticle state is exactly 1 
and this state does not contribute to the pairing correlations.   
In EFA~\cite{Perez-Martin:2008,Schunck:2023,Pore:2024}  
the occupation probability of the ``flipped quasiparticle state'' is $1-n_k$,
which includes both time-reversal single-particle states,
and its contribution to the pairing gap is
$\propto \sqrt{ n_k(1-n_k)}$. 
EFA in addition does not break time-reversal symmetry, 
the particle number parity is arguably +1 instead of -1, 
and under a full spatial rotation the many-body 
wave function does not acquire the phase $\exp(i\pi)=-1$, 
all of these aspects being expected in any system with an odd number of fermions. 
Unlike for other broken symmetries in mean-field frameworks, 
such as center-of-mass translational invariance, spatial rotation, spatial parity, 
time-reversal and gauge parities,  it is not clear if these symmetries 
can be restored and what effects on the nuclear spectra and many-body wave functions are lost.

Figs. 4 and 5 in the main text  illustrate 
the significant role of time-odd mean field terms, always present 
in time-dependent runs~\cite{Bender:2003,Shi:2021} for both 
even-even and odd nuclei. With the exception of the initially Pauli blocked state 
in odd-nucleon systems all quasiparticle states appear predominantly as almost doublets during the descent 
from outer saddle to scission in odd-nucleon systems, which is the longest in time 
part of the trajectory in our 
simulation, mainly due to the relatively slow nuclear collective motion down the ```hill.'' In even-even 
nuclei these differences between the partners of a ``doublet'' are relatively small,  but they are almost an 
order of magnitude larger 
in odd-nucleon systems.  Since the ``extra'' nucleon
contribution to any densities is about $1/N$ and/or $1/Z$ in such nuclei it is expected that these can global 
effects of comparable magnitude in such nuclei.  Such a dramatic difference between the occupation 
probabilities evaluated either from the time-dependent quasiparticle wave functions, or from the set of 
canonical wave functions determined at each time step, was observed recently in another superfluid system. In 
Ref.~\cite{Bulgac:2024} some of the present authors studied the time evolution of twelve quantized vortices 
in a unitary Fermi gas (UFG). Such a UFG state is highly excited and as conjectured by 
\textcite{Feynman:1955} a quantum superfluid develops turbulence in the absence of genuine dissipation, and 
after many crossings and reconnections the vortices fully annihilate and the entire system approaches thermal 
equilibrium at a finite temperature, different from the temperature of the initial state, as in case of 
induced fission. Although the occupation probabilities determined from the time-dependent quasiparticle wave 
functions show a dramatic evolution in time, the initial and final canonical occupation probabilities are 
surprisingly very similar~\cite{Bulgac:2024}, the UFG system is superfluid in both cases, even though the 
structure of the many-body wave functions at these different times is vastly different, similar to what we 
observe in case of induced fission in this study. 

In Figs. \ref{fig:trbreaking} and \ref{fig:sdens} we present additional view points on the important 
role of time reversal symmetry breaking terms in the dynamical evolution of both E-E and O-E systems, 
in the case of $^{238}$U(n,f).

\begin{table*}[ht]
    \centering
    \caption{Properties of the pre-neutron emission fission fragment yields in CGMF.}
    \begin{tabular}{rrrrrr}
     \hline\hline
        Reaction & TKE&  $Z_L$ & $A_L$ & $Z_H$ & $A_H$ \\
\hline
        $^{235}$U(n,f)& 172.07(9.03) & 38.31(2.21) & 97.07(5.66) & 53.69(2.21) & 138.93(5.66)  \\
         $^{237}$Np(n,f)  &174.31(9.66)& 38.87(2.46) & 98.36(6.31) & 54.13(2.46) & 139.64(6.31) \\ 
        $^{238}$U(n,f) &171.21(9.49)& 38.56(2.34) & 99.12(6.16) & 53.44(2.34) &139.88(6.16)  \\
 $^{239}$Pu(n,f) &177.18(9.42)& 39.56(2.61) & 99.86(6.67) & 54.44(2.61) & 140.14(6.67)  \\
 $^{240}$Pu(n,f) &176.13(9.40)& 39.51(2.49) & 100.16(6.33) & 54.49(2.49) &  140.84(6.33)  \\ 
 $^{242}$Pu(n,f) & 176.93(9.41)& 39.94(2.43) & 102.13(6.33) &54.06(2.43) &140.87(6.33)  \\ 
        \hline \hline
    \end{tabular}
    \label{tab:cgmf-properties}
\end{table*}

 \section{Fission Times and TKE} 

 In Fig~\ref{fig:0000}, the TKE for all nuclei is shown as a function of the initial $Q_{30}$ of the compound system at the saddle point.  Two clusters of trajectories are observed, those with $Q_{30} \gtrapprox 10$ b$^{3/2}$ and those with $Q_{30} \lessapprox 10$ b$^{3/2}$.  Trajectories with larger initial $Q_{30}$ values have a substantially higher TKE relative to trajectories with lower initial $Q_{30}$ values, which is a trend that holds at all compound excitation energies.    As shown in Fig. 3 of the main manuscript, with further discussions contained there, trajectories with lower TKE often take substantially longer to evolve from the saddle to fission, wandering on the potential energy surface for an extended time.  Note, currently all trajectories are weighed equally, which should be improved upon in the future, as the relative weights of these lower TKE trajectories can greatly affect theoretical predictions.  This is especially true at higher compound excitation energies where the more symmetric modes, those with smaller $Q_{30}$, are expected to provide a increasingly important contribution to fission observables.

 As shown in Fig.~\ref{fig:000}, $^{238}$Np exhibits a longer saddle to scission time on average when treated as a odd-odd system versus a even-even system, similar to what was observed for $^{239}$U and $^{241,243}$Pu (see Fig. 2 of the main text).  In this case however, the correlation is much weaker, as $^{238}$Np requires two quasi-particle excitations, one for neutrons and one for protons.  This means even if the quasi-particle energy of the unpaired nucleon is small in one sector it is not guaranteed to be small in the other, and, in fact, often isn't as shown in the run tables below.

\section{Fission yields in CGMF}

In Table \ref{tab:cgmf-properties} we summarize the properties of the FF distributions that are used in CGMF \cite{TALOU2021108087} to model event-by-event the emission of prompt neutrons and gamma rays. These are parameterized on the basis of experimental data and available reaction systematics. Note that in CGMF one allows an overall shift (up or down) of TKE from the experimental data in order to match the measured average number of prompt neutrons emitted in fission $\bar \nu$. As not enough data exist on the width of TKE distribution (as a function of the preneutron emission fragment mass), the standard deviation is also often adjusted to reproduce the measured or systematic neutron multiplicity distribution.

\section{Details of all runs} \label{ap:B}

In the following tables, following these clarifications,  
we provide details for all the runs, for both the ``seed'' nuclei and the nuclei with an odd number of nucleons.

\begin{itemize}
\item $^{239}$U, 6 even-even runs for ``seed'' nuclei and 23 odd runs,
\item $^{241}$Pu, 13 even-even runs for ``seed'' and 43 odd runs,
\item $^{243}$Pu, 10 even-even runs for ``seed'' and 35 odd runs,
\item $^{238}$Np, 11 even-even runs for ``seed'' and 48 odd runs.
\end{itemize}

In the first column ``Flip'' of each table we identify the quasiparticle number of the flipped state, obtained for the ``seed'' nucleus,  
in the case of nuclei with an odd nucleon number or with symbol ``EE'' the results of the run for the corresponding ``seed'' nucleus. 
In next 3 columns we report the total initial energy in MeV, $N$, and $Z$. 

In the following 4 columns we report the flipped quasiparticle energy, its occupation probability, the average and exact value of the 
angular momentum $j_z$,  of the corresponding state. Since quasiparticle states appear as doublets in the case of even-even nuclei, 
the quasiparticle wave function is typically an arbitrary linear combination of the two time-reversed states $k$ and $\overline{k}$ and 
the ``exact value of $j_z$ is obtained by diagonalizing the operator $\hat{j}_z$ in the 2-by-2 space of the doublet. 
Unfortunately, on a discretized lattice the orbital angular momentum is never an integer as expected in theory, a deficiency of all 
lattice approaches in condensed matter, nuclear physics, and QCD for example. 

In the following two columns we report the initial average values of multipole operators  $\hat{Q}_{20},\,\hat{Q}_{30}$ in barns and barns$^{3/2}$ respectively.

In the column TKE we report the corresponding total kinetic energy  of the FFs in MeV.
 In the column $t_{\mathrm{scission}}$ we report the scission times in fm/c.
 
 In the last seven columns we report the evaluated FF $Z$, $N$, and intrinsic energies of the FFs, and in the error in the total energy
 of the entire system $\Delta E = E(t_{final})-E(t_{initial})$.
 
 The symmetric and asymmetric runs are defined by the value of the initial octupole deformation of the compound nucleus, 
 symmetric being those with $Q_{30}< 2.5$ barns$^{3/2}$.

\clearpage
\begin{sidewaystable}[h]
\vspace{3.5in}
%\begin{table}[h]
\centering
\begin{tabular}{|c|c|c|c|c|c|c|c|c|c|c|c|c|c|c|c|c|c|c|}
\hline
Flip & $E$ & $Z$ & $N$ & $E_\text{QP}$ & $n_k$ & $\langle j_z \rangle$ & $j_z$ & $Q_{20}$ & $Q_{30}$ & TKE & $t_\text{scission}$ & $Z_\text{HFF}$ & $Z_\text{LFF}$ & $N_\text{HFF}$ & $N_\text{LFF}$ & $E^\text{intr}_\text{HFF}$ & $E^\text{intr}_\text{LFF}$ & $\Delta E$ \\ \hline 
4 & -1813.08 & 94.00 & 147.02 & 1.00 & 0.49 & -0.00 & -0.00 & 137.66 & 13.21 & 180.06 & 2210 & 52.50 & 41.50 & 83.16 & 63.86 & -1121.31 & -871.87 & -0.04\\ \hline
3603 & -1808.67 & 94.00 & 146.97 & 5.11 & 0.51 & -0.11 & -0.50 & 137.80 & 13.20 & 177.26 & 1849 & 53.10 & 40.90 & 83.59 & 63.38 & -1127.44 & -858.53 & -0.05\\ \hline
EE & -1813.97 & 94.00 & 147.00 & - & - & - & - & 137.65 & 13.21 & 174.46 & 1874 & 53.35 & 40.65 & 84.10 & 62.90 & -1134.48 & -854.00 & -0.05\\ \hline
\end{tabular}
\caption{$^{241}$Pu Run 00 Asymmetric}
%\end{table}
%\begin{table}[h]
\centering
\begin{tabular}{|c|c|c|c|c|c|c|c|c|c|c|c|c|c|c|c|c|c|c|}
\hline
Flip & $E$ & $Z$ & $N$ & $E_\text{QP}$ & $n_k$ & $\langle j_z \rangle$ & $j_z$ & $Q_{20}$ & $Q_{30}$ & TKE & $t_\text{scission}$ & $Z_\text{HFF}$ & $Z_\text{LFF}$ & $N_\text{HFF}$ & $N_\text{LFF}$ & $E^\text{intr}_\text{HFF}$ & $E^\text{intr}_\text{LFF}$ & $\Delta E$ \\ \hline 
0 & -1812.82 & 94.00 & 146.79 & 0.94 & 0.61 & -0.64 & -0.65 & 200.43 & 30.92 & 175.95 & 1293 & 52.67 & 41.33 & 82.95 & 63.84 & -1120.38 & -868.44 & -0.05\\ \hline
1811 & -1809.67 & 94.00 & 146.65 & 3.29 & 0.68 & 0.14 & 0.33 & 200.15 & 30.83 & 171.15 & 1461 & 53.42 & 40.58 & 84.22 & 62.42 & -1129.28 & -851.58 & -0.03\\ \hline
3614 & -1809.71 & 94.00 & 147.03 & 5.34 & 0.48 & 0.54 & 0.54 & 201.08 & 30.75 & 177.14 & 1254 & 52.29 & 41.71 & 82.74 & 64.29 & -1113.08 & -873.80 & -0.04\\ \hline
10800 & -1805.47 & 94.00 & 146.85 & 8.61 & 0.57 & 2.04 & 2.08 & 200.94 & 30.60 & 174.62 & 1299 & 52.90 & 41.10 & 83.74 & 63.11 & -1125.15 & -854.97 & -0.04\\ \hline
EE & -1814.89 & 94.00 & 147.00 & - & - & - & - & 200.62 & 30.85 & 176.82 & 1319 & 52.64 & 41.36 & 83.33 & 63.67 & -1123.65 & -868.11 & -0.05\\ \hline
\end{tabular}
\caption{$^{241}$Pu Run 01 Asymmetric}
%\end{table}
%\begin{table}[h]
\centering
\begin{tabular}{|c|c|c|c|c|c|c|c|c|c|c|c|c|c|c|c|c|c|c|}
\hline
Flip & $E$ & $Z$ & $N$ & $E_\text{QP}$ & $n_k$ & $\langle j_z \rangle$ & $j_z$ & $Q_{20}$ & $Q_{30}$ & TKE & $t_\text{scission}$ & $Z_\text{HFF}$ & $Z_\text{LFF}$ & $N_\text{HFF}$ & $N_\text{LFF}$ & $E^\text{intr}_\text{HFF}$ & $E^\text{intr}_\text{LFF}$ & $\Delta E$ \\ \hline 
0 & -1815.85 & 94.00 & 147.29 & 0.73 & 0.35 & -1.24 & -1.24 & 183.76 & 26.00 & 173.17 & 1622 & 52.45 & 41.55 & 83.06 & 64.23 & -1114.83 & -874.21 & -0.03\\ \hline
2 & -1815.71 & 94.00 & 147.28 & 0.78 & 0.36 & -1.04 & -2.70 & 183.70 & 26.01 & 169.49 & 1713 & 53.11 & 40.89 & 84.04 & 63.24 & -1125.07 & -860.16 & -0.03\\ \hline
6 & -1811.83 & 94.00 & 146.61 & 1.02 & 0.69 & 0.09 & 0.09 & 182.95 & 25.97 & 174.42 & 1700 & 52.56 & 41.44 & 82.96 & 63.65 & -1117.36 & -868.93 & -0.04\\ \hline
1836 & -1809.98 & 94.00 & 146.91 & 4.50 & 0.54 & -1.88 & -1.91 & 183.24 & 26.14 & 173.55 & 1461 & 53.05 & 40.95 & 83.70 & 63.21 & -1120.81 & -862.75 & -0.04\\ \hline
EE & -1814.97 & 94.00 & 147.00 & - & - & - & - & 183.49 & 26.03 & 172.81 & 1480 & 53.12 & 40.88 & 83.91 & 63.09 & -1126.01 & -861.80 & -0.03\\ \hline
\end{tabular}
\caption{$^{241}$Pu Run 02 Asymmetric}
%\end{table}
%\begin{table}[h]
\centering
\begin{tabular}{|c|c|c|c|c|c|c|c|c|c|c|c|c|c|c|c|c|c|c|}
\hline
Flip & $E$ & $Z$ & $N$ & $E_\text{QP}$ & $n_k$ & $\langle j_z \rangle$ & $j_z$ & $Q_{20}$ & $Q_{30}$ & TKE & $t_\text{scission}$ & $Z_\text{HFF}$ & $Z_\text{LFF}$ & $N_\text{HFF}$ & $N_\text{LFF}$ & $E^\text{intr}_\text{HFF}$ & $E^\text{intr}_\text{LFF}$ & $\Delta E$ \\ \hline 
0 & -1809.94 & 94.00 & 146.73 & 0.65 & 0.63 & -1.98 & -1.98 & 158.06 & 19.53 & 175.59 & 1849 & 53.36 & 40.64 & 83.64 & 63.09 & -1130.84 & -854.64 & 0.05\\ \hline
4 & -1811.77 & 94.00 & 147.06 & 0.61 & 0.47 & 1.06 & 1.06 & 158.34 & 19.56 & 170.12 & 1971 & 53.47 & 40.53 & 84.33 & 62.73 & -1132.72 & -849.12 & 0.05\\ \hline
1826 & -1807.94 & 94.00 & 147.06 & 4.41 & 0.47 & -1.89 & -1.89 & 158.26 & 19.63 & 178.71 & 1797 & 52.96 & 41.04 & 83.81 & 63.25 & -1125.47 & -861.14 & 0.04\\ \hline
EE & -1812.04 & 94.00 & 147.00 & - & - & - & - & 158.25 & 19.56 & 178.02 & 1732 & 52.74 & 41.27 & 83.03 & 63.97 & -1121.88 & -868.15 & 0.03\\ \hline
\end{tabular}
\caption{$^{241}$Pu Run 03 Asymmetric}
%\end{table}
%\begin{table}[h]
\centering
\begin{tabular}{|c|c|c|c|c|c|c|c|c|c|c|c|c|c|c|c|c|c|c|}
\hline
Flip & $E$ & $Z$ & $N$ & $E_\text{QP}$ & $n_k$ & $\langle j_z \rangle$ & $j_z$ & $Q_{20}$ & $Q_{30}$ & TKE & $t_\text{scission}$ & $Z_\text{HFF}$ & $Z_\text{LFF}$ & $N_\text{HFF}$ & $N_\text{LFF}$ & $E^\text{intr}_\text{HFF}$ & $E^\text{intr}_\text{LFF}$ & $\Delta E$ \\ \hline 
0 & -1814.78 & 94.00 & 147.11 & 0.79 & 0.44 & -0.70 & -0.70 & 207.39 & 32.83 & 172.38 & 1331 & 53.01 & 40.99 & 83.92 & 63.20 & -1127.40 & -859.80 & -0.04\\ \hline
2 & -1812.96 & 94.00 & 146.79 & 0.92 & 0.60 & 0.99 & 1.13 & 206.96 & 32.87 & 173.32 & 1286 & 52.84 & 41.16 & 83.69 & 63.11 & -1126.00 & -860.33 & -0.05\\ \hline
EE & -1814.96 & 94.00 & 147.00 & - & - & - & - & 207.21 & 32.79 & 173.38 & 1241 & 52.83 & 41.17 & 83.61 & 63.39 & -1125.30 & -863.09 & -0.04\\ \hline
\end{tabular}
\caption{$^{241}$Pu Run 04 Asymmetric}
%\end{table}

\end{sidewaystable}
\clearpage

\clearpage
\begin{sidewaystable}[h]
\vspace{3.5in}
%\begin{table}[h]
\centering
\begin{tabular}{|c|c|c|c|c|c|c|c|c|c|c|c|c|c|c|c|c|c|c|}
\hline
Flip & $E$ & $Z$ & $N$ & $E_\text{QP}$ & $n_k$ & $\langle j_z \rangle$ & $j_z$ & $Q_{20}$ & $Q_{30}$ & TKE & $t_\text{scission}$ & $Z_\text{HFF}$ & $Z_\text{LFF}$ & $N_\text{HFF}$ & $N_\text{LFF}$ & $E^\text{intr}_\text{HFF}$ & $E^\text{intr}_\text{LFF}$ & $\Delta E$ \\ \hline 
0 & -1813.43 & 94.00 & 146.97 & 0.90 & 0.51 & -0.64 & -1.74 & 144.27 & 15.01 & 177.32 & 1829 & 53.19 & 40.81 & 83.91 & 63.06 & -1131.25 & -859.56 & -0.06\\ \hline
4 & -1812.23 & 94.00 & 146.78 & 1.04 & 0.61 & 0.16 & 0.16 & 144.13 & 15.00 & 174.73 & 1965 & 53.25 & 40.75 & 83.90 & 62.88 & -1130.82 & -856.19 & -0.05\\ \hline
6 & -1813.69 & 94.00 & 147.04 & 1.01 & 0.48 & 0.07 & 1.91 & 144.37 & 15.01 & 177.80 & 2004 & 52.98 & 41.02 & 83.51 & 63.53 & -1126.30 & -865.25 & -0.06\\ \hline
1830 & -1809.44 & 94.00 & 146.94 & 4.65 & 0.53 & 0.67 & 0.76 & 144.01 & 15.05 & 173.45 & 1861 & 53.17 & 40.83 & 83.84 & 63.10 & -1129.02 & -853.91 & -0.05\\ \hline
EE & -1814.46 & 94.00 & 147.00 & - & - & - & - & 144.33 & 15.01 & 176.06 & 1771 & 53.10 & 40.90 & 83.83 & 63.17 & -1131.24 & -859.33 & -0.05\\ \hline
\end{tabular}
\caption{$^{241}$Pu Run 05 Asymmetric}
%\end{table}
%\begin{table}[h]
\centering
\begin{tabular}{|c|c|c|c|c|c|c|c|c|c|c|c|c|c|c|c|c|c|c|}
\hline
Flip & $E$ & $Z$ & $N$ & $E_\text{QP}$ & $n_k$ & $\langle j_z \rangle$ & $j_z$ & $Q_{20}$ & $Q_{30}$ & TKE & $t_\text{scission}$ & $Z_\text{HFF}$ & $Z_\text{LFF}$ & $N_\text{HFF}$ & $N_\text{LFF}$ & $E^\text{intr}_\text{HFF}$ & $E^\text{intr}_\text{LFF}$ & $\Delta E$ \\ \hline 
0 & -1814.31 & 94.00 & 147.19 & 0.91 & 0.41 & -1.90 & -1.94 & 139.48 & 9.43 & 167.88 & 2101 & 54.00 & 40.00 & 85.62 & 61.57 & -1144.62 & -837.61 & -0.03\\ \hline
2 & -1814.34 & 94.00 & 147.20 & 0.96 & 0.40 & 2.44 & 2.80 & 139.61 & 9.45 & 182.01 & 2146 & 52.44 & 41.56 & 83.08 & 64.12 & -1120.96 & -875.44 & -0.06\\ \hline
4 & -1812.82 & 94.00 & 146.94 & 1.03 & 0.53 & -0.05 & -0.05 & 139.41 & 9.44 & 177.14 & 3988 & 52.45 & 41.55 & 83.04 & 63.90 & -1118.29 & -871.72 & -0.05\\ \hline
6 & -1813.45 & 94.00 & 147.05 & 1.03 & 0.47 & -1.17 & -2.00 & 139.52 & 9.44 & 175.93 & 2372 & 52.89 & 41.12 & 83.63 & 63.42 & -1126.27 & -863.19 & -0.08\\ \hline
30 & -1812.16 & 94.00 & 147.06 & 2.32 & 0.47 & -0.06 & -0.08 & 139.43 & 9.42 & 171.79 & 1978 & 53.32 & 40.68 & 83.93 & 63.13 & -1129.22 & -854.77 & -0.03\\ \hline
32 & -1811.56 & 94.00 & 146.96 & 2.34 & 0.52 & 0.21 & 0.23 & 139.30 & 9.42 & 173.54 & 2004 & 53.01 & 40.99 & 83.66 & 63.30 & -1124.64 & -860.49 & -0.03\\ \hline
EE & -1814.18 & 94.00 & 147.00 & - & - & - & - & 139.47 & 9.44 & 173.54 & 1816 & 53.33 & 40.67 & 83.94 & 63.06 & -1132.59 & -855.17 & -0.04\\ \hline
\end{tabular}
\caption{$^{241}$Pu Run 06 Asymmetric}
%\end{table}
%\begin{table}[h]
\centering
\begin{tabular}{|c|c|c|c|c|c|c|c|c|c|c|c|c|c|c|c|c|c|c|}
\hline
Flip & $E$ & $Z$ & $N$ & $E_\text{QP}$ & $n_k$ & $\langle j_z \rangle$ & $j_z$ & $Q_{20}$ & $Q_{30}$ & TKE & $t_\text{scission}$ & $Z_\text{HFF}$ & $Z_\text{LFF}$ & $N_\text{HFF}$ & $N_\text{LFF}$ & $E^\text{intr}_\text{HFF}$ & $E^\text{intr}_\text{LFF}$ & $\Delta E$ \\ \hline 
0 & -1814.44 & 94.03 & 147.11 & 0.89 & 0.44 & 1.00 & 2.68 & 164.12 & 7.77 & 175.95 & 2631 & 52.82 & 41.21 & 83.27 & 63.84 & -1122.95 & -867.51 & -0.06\\ \hline
2 & -1813.03 & 94.03 & 146.88 & 0.96 & 0.56 & -2.14 & -2.29 & 163.91 & 7.76 & 175.71 & 1900 & 53.12 & 40.90 & 83.93 & 62.95 & -1129.98 & -858.80 & -0.04\\ \hline
3 & -1813.03 & 94.03 & 146.88 & 0.96 & 0.56 & 2.14 & 2.29 & 163.91 & 7.76 & 175.71 & 1900 & 53.12 & 40.90 & 83.93 & 62.95 & -1129.98 & -858.80 & -0.04\\ \hline
7222 & -1806.88 & 94.03 & 147.00 & 7.75 & 0.50 & 0.32 & 0.33 & 165.54 & 7.35 & 176.74 & 1887 & 52.57 & 41.46 & 83.44 & 63.56 & -1118.72 & -864.95 & -0.05\\ \hline
EE & -1814.68 & 94.03 & 147.00 & - & - & - & - & 164.03 & 7.77 & 176.57 & 1719 & 52.92 & 41.10 & 83.34 & 63.66 & -1124.35 & -866.94 & -0.04\\ \hline
\end{tabular}
\caption{$^{241}$Pu Run 07 Asymmetric}
%\end{table}

%\begin{table}[h]
\centering
\begin{tabular}{|c|c|c|c|c|c|c|c|c|c|c|c|c|c|c|c|c|c|c|}
\hline
Flip & $E$ & $Z$ & $N$ & $E_\text{QP}$ & $n_k$ & $\langle j_z \rangle$ & $j_z$ & $Q_{20}$ & $Q_{30}$ & TKE & $t_\text{scission}$ & $Z_\text{HFF}$ & $Z_\text{LFF}$ & $N_\text{HFF}$ & $N_\text{LFF}$ & $E^\text{intr}_\text{HFF}$ & $E^\text{intr}_\text{LFF}$ & $\Delta E$ \\ \hline 
0 & -1814.11 & 94.00 & 147.02 & 0.92 & 0.49 & -1.27 & -2.65 & 174.67 & 7.24 & 174.33 & 2133 & 51.55 & 42.45 & 81.81 & 65.21 & -1099.66 & -888.82 & -0.05\\ \hline
2 & -1812.16 & 94.00 & 146.71 & 1.07 & 0.65 & -2.23 & -2.26 & 174.54 & 7.13 & 165.82 & 1836 & 53.21 & 40.79 & 83.63 & 63.08 & -1125.13 & -852.88 & -0.03\\ \hline
4 & -1811.96 & 94.00 & 146.66 & 1.02 & 0.67 & -1.70 & -2.01 & 174.32 & 7.23 & 174.61 & 1752 & 52.88 & 41.12 & 83.65 & 63.01 & -1124.09 & -862.53 & -0.05\\ \hline
EE & -1814.91 & 94.00 & 147.00 & - & - & - & - & 174.65 & 7.24 & 170.50 & 1745 & 53.22 & 40.78 & 83.73 & 63.27 & -1125.75 & -859.70 & -0.04\\ \hline
\end{tabular}
\caption{$^{241}$Pu Run 08 Asymmetric}
%\end{table}

\end{sidewaystable}

\clearpage
\begin{sidewaystable}[h]
\vspace{3.5in}
%\begin{table}[h]
\centering
\begin{tabular}{|c|c|c|c|c|c|c|c|c|c|c|c|c|c|c|c|c|c|c|}
\hline
Flip & $E$ & $Z$ & $N$ & $E_\text{QP}$ & $n_k$ & $\langle j_z \rangle$ & $j_z$ & $Q_{20}$ & $Q_{30}$ & TKE & $t_\text{scission}$ & $Z_\text{HFF}$ & $Z_\text{LFF}$ & $N_\text{HFF}$ & $N_\text{LFF}$ & $E^\text{intr}_\text{HFF}$ & $E^\text{intr}_\text{LFF}$ & $\Delta E$ \\ \hline 
0 & -1813.31 & 94.00 & 146.86 & 0.95 & 0.57 & 0.05 & 2.61 & 188.07 & 6.36 & 171.68 & 1771 & 50.37 & 43.63 & 79.71 & 67.15 & -1073.49 & -911.54 & -0.04\\ \hline
4 & -1814.97 & 94.00 & 147.16 & 0.99 & 0.42 & -0.23 & -0.38 & 188.23 & 6.43 & 172.97 & 1590 & 51.72 & 42.28 & 82.65 & 64.51 & -1103.99 & -883.98 & -0.04\\ \hline
10 & -1812.76 & 94.00 & 146.81 & 1.21 & 0.60 & 3.17 & 4.44 & 188.11 & 6.28 & 158.69 & 1596 & 50.83 & 43.17 & 80.46 & 66.34 & -1069.80 & -901.64 & 0.01\\ \hline
20 & -1814.00 & 94.00 & 147.21 & 1.63 & 0.67 & 0.02 & 1.11 & 188.34 & 6.24 & 169.08 & 1545 & 51.10 & 42.90 & 81.35 & 65.86 & -1086.25 & -896.85 & -0.02\\ \hline
EE & -1815.03 & 94.00 & 147.00 & - & - & - & - & 188.19 & 6.36 & 167.64 & 1519 & 50.66 & 43.34 & 80.78 & 66.22 & -1079.80 & -902.91 & -0.03\\ \hline
\end{tabular}
\caption{$^{241}$Pu Run 09 Asymmetric}
%\end{table}
%\begin{table}[h]
\centering
\begin{tabular}{|c|c|c|c|c|c|c|c|c|c|c|c|c|c|c|c|c|c|c|}
\hline
Flip & $E$ & $Z$ & $N$ & $E_\text{QP}$ & $n_k$ & $\langle j_z \rangle$ & $j_z$ & $Q_{20}$ & $Q_{30}$ & TKE & $t_\text{scission}$ & $Z_\text{HFF}$ & $Z_\text{LFF}$ & $N_\text{HFF}$ & $N_\text{LFF}$ & $E^\text{intr}_\text{HFF}$ & $E^\text{intr}_\text{LFF}$ & $\Delta E$ \\ \hline 
0 & -1815.04 & 93.98 & 147.07 & 0.80 & 0.46 & 0.64 & 0.64 & 204.37 & 2.37 & 149.37 & 10348 & 51.73 & 42.25 & 81.99 & 65.08 & -1080.97 & -883.24 & 0.01\\ \hline
4 & -1813.12 & 93.98 & 146.74 & 0.93 & 0.63 & -1.03 & -1.04 & 203.96 & 2.37 & 157.97 & 6457 & 50.47 & 43.50 & 80.22 & 66.52 & -1062.14 & -908.94 & 0.01\\ \hline
16210 & -1805.99 & 93.98 & 147.15 & 10.25 & 0.43 & 0.74 & 1.61 & 207.28 & 2.14 & 151.15 & 4867 & 51.43 & 42.55 & 81.45 & 65.70 & -1067.79 & -889.37 & -0.01\\ \hline
EE & -1815.45 & 93.98 & 147.00 & - & - & - & - & 204.30 & 2.37 & 149.07 & 4802 & 50.43 & 43.55 & 79.37 & 67.63 & -1051.53 & -912.99 & -0.00\\ \hline
\end{tabular}
\caption{$^{241}$Pu Run 00 Symmetric}
%\end{table}
%\begin{table}[h]
\centering
\begin{tabular}{|c|c|c|c|c|c|c|c|c|c|c|c|c|c|c|c|c|c|c|}
\hline
Flip & $E$ & $Z$ & $N$ & $E_\text{QP}$ & $n_k$ & $\langle j_z \rangle$ & $j_z$ & $Q_{20}$ & $Q_{30}$ & TKE & $t_\text{scission}$ & $Z_\text{HFF}$ & $Z_\text{LFF}$ & $N_\text{HFF}$ & $N_\text{LFF}$ & $E^\text{intr}_\text{HFF}$ & $E^\text{intr}_\text{LFF}$ & $\Delta E$ \\ \hline 
0 & -1810.85 & 93.99 & 146.82 & 0.51 & 0.59 & 0.64 & 0.64 & 200.72 & 0.61 & 149.02 & 10102 & 50.99 & 42.99 & 81.07 & 65.75 & -1063.15 & -896.43 & 0.30\\ \hline
2 & -1809.81 & 93.99 & 146.64 & 0.61 & 0.68 & 1.01 & 1.01 & 200.42 & 0.61 & 155.79 & 11847 & 49.38 & 44.61 & 78.27 & 68.37 & -1035.52 & -929.09 & 0.98\\ \hline
EE & -1812.26 & 93.99 & 147.00 & - & - & - & - & 200.89 & 0.61 & 149.74 & 4440 & 51.42 & 42.57 & 81.31 & 65.69 & -1071.69 & -890.17 & 0.14\\ \hline
\end{tabular}
\caption{$^{241}$Pu Run 01 Symmetric}
%\end{table}
%\begin{table}[h]
\centering
\begin{tabular}{|c|c|c|c|c|c|c|c|c|c|c|c|c|c|c|c|c|c|c|}
\hline
Flip & $E$ & $Z$ & $N$ & $E_\text{QP}$ & $n_k$ & $\langle j_z \rangle$ & $j_z$ & $Q_{20}$ & $Q_{30}$ & TKE & $t_\text{scission}$ & $Z_\text{HFF}$ & $Z_\text{LFF}$ & $N_\text{HFF}$ & $N_\text{LFF}$ & $E^\text{intr}_\text{HFF}$ & $E^\text{intr}_\text{LFF}$ & $\Delta E$ \\ \hline 
0 & -1812.06 & 93.69 & 147.01 & 0.71 & 0.50 & -0.64 & -0.64 & 204.78 & 1.30 & 147.24 & 11608 & 50.95 & 42.74 & 80.40 & 66.60 & -1060.29 & -898.93 & 0.09\\ \hline
1828 & -1807.99 & 93.69 & 146.91 & 4.20 & 0.55 & 0.08 & 1.33 & 204.90 & 1.30 & 150.51 & 4498 & 49.98 & 43.71 & 79.47 & 67.44 & -1044.88 & -913.60 & 0.02\\ \hline
EE & -1812.72 & 93.69 & 147.00 & - & - & - & - & 204.77 & 1.30 & 154.98 & 5151 & 51.16 & 42.53 & 80.74 & 66.26 & -1070.79 & -896.88 & 0.04\\ \hline
\end{tabular}
\caption{$^{241}$Pu Run 02 Symmetric}
%\end{table}
%\begin{table}[h]
\centering
\begin{tabular}{|c|c|c|c|c|c|c|c|c|c|c|c|c|c|c|c|c|c|c|}
\hline
Flip & $E$ & $Z$ & $N$ & $E_\text{QP}$ & $n_k$ & $\langle j_z \rangle$ & $j_z$ & $Q_{20}$ & $Q_{30}$ & TKE & $t_\text{scission}$ & $Z_\text{HFF}$ & $Z_\text{LFF}$ & $N_\text{HFF}$ & $N_\text{LFF}$ & $E^\text{intr}_\text{HFF}$ & $E^\text{intr}_\text{LFF}$ & $\Delta E$ \\ \hline 
0 & -1800.38 & 92.00 & 146.91 & 0.87 & 0.54 & 0.70 & 1.89 & 148.09 & 16.75 & 171.00 & 2198 & 51.71 & 40.29 & 83.47 & 63.44 & -1116.18 & -855.23 & -0.04\\ \hline
8 & -1800.82 & 92.00 & 147.01 & 0.94 & 0.49 & 0.48 & 0.76 & 148.25 & 16.76 & 175.38 & 2314 & 51.62 & 40.38 & 82.94 & 64.07 & -1114.89 & -861.35 & -0.04\\ \hline
10 & -1800.40 & 92.00 & 146.94 & 1.00 & 0.53 & -0.17 & -0.17 & 148.18 & 16.76 & 169.68 & 7834 & 51.47 & 40.53 & 83.03 & 63.92 & -1112.25 & -857.86 & -0.03\\ \hline
14426 & -1791.93 & 92.00 & 147.03 & 9.86 & 0.49 & 0.67 & 0.78 & 149.93 & 16.60 & 169.20 & 1958 & 51.84 & 40.16 & 83.59 & 63.44 & -1112.72 & -848.42 & -0.01\\ \hline
EE & -1798.93 & 92.00 & 147.00 & - & - & - & - & 145.91 & 16.67 & 170.78 & 2172 & 51.72 & 40.28 & 83.30 & 63.70 & -1115.14 & -854.53 & 0.05\\ \hline
\end{tabular}
\caption{$^{239}$U Run 08}
%\end{table}

\end{sidewaystable}

\clearpage
\begin{sidewaystable}[h]
\vspace{3.5in}
%\begin{table}[h]
\centering
\begin{tabular}{|c|c|c|c|c|c|c|c|c|c|c|c|c|c|c|c|c|c|c|}
\hline
Flip & $E$ & $Z$ & $N$ & $E_\text{QP}$ & $n_k$ & $\langle j_z \rangle$ & $j_z$ & $Q_{20}$ & $Q_{30}$ & TKE & $t_\text{scission}$ & $Z_\text{HFF}$ & $Z_\text{LFF}$ & $N_\text{HFF}$ & $N_\text{LFF}$ & $E^\text{intr}_\text{HFF}$ & $E^\text{intr}_\text{LFF}$ & $\Delta E$ \\ \hline 
0 & -1800.55 & 92.00 & 146.85 & 0.94 & 0.57 & -0.70 & -2.03 & 157.60 & 19.15 & 161.65 & 2146 & 53.14 & 38.86 & 86.18 & 60.67 & -1140.64 & -821.58 & -0.02\\ \hline
6 & -1802.18 & 92.00 & 147.17 & 0.90 & 0.42 & 1.92 & 1.93 & 157.80 & 19.14 & 164.14 & 2185 & 52.51 & 39.49 & 84.43 & 62.74 & -1126.81 & -839.53 & -0.02\\ \hline
1828 & -1797.06 & 92.00 & 146.84 & 4.35 & 0.58 & -2.68 & -2.69 & 157.64 & 19.20 & 173.67 & 1836 & 51.55 & 40.45 & 83.35 & 63.49 & -1111.92 & -858.84 & -0.04\\ \hline
1832 & -1798.71 & 92.00 & 147.17 & 4.35 & 0.41 & -2.58 & -2.64 & 157.79 & 19.19 & 172.63 & 1771 & 51.39 & 40.61 & 83.11 & 64.06 & -1109.17 & -862.20 & -0.04\\ \hline
EE & -1799.04 & 92.00 & 147.00 & - & - & - & - & 154.84 & 19.07 & 168.75 & 1926 & 51.92 & 40.08 & 83.59 & 63.41 & -1116.57 & -851.16 & 0.05\\ \hline
\end{tabular}
\caption{$^{239}$U Run 09}
%\end{table}
%\begin{table}[h]
\centering
\begin{tabular}{|c|c|c|c|c|c|c|c|c|c|c|c|c|c|c|c|c|c|c|}
\hline
Flip & $E$ & $Z$ & $N$ & $E_\text{QP}$ & $n_k$ & $\langle j_z \rangle$ & $j_z$ & $Q_{20}$ & $Q_{30}$ & TKE & $t_\text{scission}$ & $Z_\text{HFF}$ & $Z_\text{LFF}$ & $N_\text{HFF}$ & $N_\text{LFF}$ & $E^\text{intr}_\text{HFF}$ & $E^\text{intr}_\text{LFF}$ & $\Delta E$ \\ \hline 
6 & -1800.77 & 92.00 & 147.02 & 1.08 & 0.49 & -1.04 & -2.01 & 139.75 & 10.03 & 170.67 & 2734 & 51.90 & 40.10 & 83.59 & 63.43 & -1119.21 & -852.27 & -0.04\\ \hline
3604 & -1796.91 & 92.00 & 147.03 & 4.92 & 0.49 & -0.12 & -0.13 & 139.72 & 10.04 & 169.83 & 2120 & 51.98 & 40.02 & 83.73 & 63.29 & -1118.62 & -848.17 & -0.04\\ \hline
12616 & -1792.75 & 92.00 & 147.07 & 9.31 & 0.46 & 0.50 & 0.73 & 140.89 & 9.88 & 167.54 & 1913 & 51.68 & 40.32 & 82.97 & 64.10 & -1109.48 & -850.82 & -0.01\\ \hline
12618 & -1792.22 & 92.00 & 146.98 & 9.38 & 0.51 & 2.29 & 2.54 & 140.36 & 10.06 & 166.98 & 2036 & 52.04 & 39.96 & 83.99 & 62.99 & -1113.07 & -846.18 & -0.04\\ \hline
EE & -1801.72 & 92.00 & 147.00 & - & - & - & - & 139.72 & 10.03 & 170.56 & 1939 & 51.53 & 40.47 & 83.10 & 63.90 & -1114.15 & -858.17 & -0.04\\ \hline
\end{tabular}
\caption{$^{239}$U Run 11}
%\end{table}
%\begin{table}[h]
\centering
\begin{tabular}{|c|c|c|c|c|c|c|c|c|c|c|c|c|c|c|c|c|c|c|}
\hline
Flip & $E$ & $Z$ & $N$ & $E_\text{QP}$ & $n_k$ & $\langle j_z \rangle$ & $j_z$ & $Q_{20}$ & $Q_{30}$ & TKE & $t_\text{scission}$ & $Z_\text{HFF}$ & $Z_\text{LFF}$ & $N_\text{HFF}$ & $N_\text{LFF}$ & $E^\text{intr}_\text{HFF}$ & $E^\text{intr}_\text{LFF}$ & $\Delta E$ \\ \hline 
0 & -1801.85 & 92.00 & 147.18 & 0.90 & 0.41 & -1.23 & -2.64 & 169.01 & 7.76 & 170.92 & 2178 & 51.17 & 40.83 & 82.99 & 64.19 & -1104.50 & -868.31 & -0.04\\ \hline
2 & -1799.96 & 92.00 & 146.84 & 0.94 & 0.58 & -1.73 & -2.13 & 168.70 & 7.76 & 169.13 & 1939 & 51.41 & 40.59 & 82.93 & 63.91 & -1107.67 & -861.46 & -0.04\\ \hline
12630 & -1791.74 & 92.00 & 147.00 & 9.94 & 0.50 & 0.00 & 0.02 & 170.01 & 7.85 & 167.97 & 1771 & 51.12 & 40.88 & 82.00 & 64.99 & -1096.15 & -863.57 & -0.01\\ \hline
EE & -1801.73 & 92.00 & 147.00 & - & - & - & - & 168.86 & 7.76 & 167.70 & 1797 & 50.95 & 41.05 & 82.20 & 64.80 & -1099.42 & -870.04 & -0.03\\ \hline
\end{tabular}
\caption{$^{239}$U Run 12}
%\end{table}
%\begin{table}[h]
\centering
\begin{tabular}{|c|c|c|c|c|c|c|c|c|c|c|c|c|c|c|c|c|c|c|}
\hline
Flip & $E$ & $Z$ & $N$ & $E_\text{QP}$ & $n_k$ & $\langle j_z \rangle$ & $j_z$ & $Q_{20}$ & $Q_{30}$ & TKE & $t_\text{scission}$ & $Z_\text{HFF}$ & $Z_\text{LFF}$ & $N_\text{HFF}$ & $N_\text{LFF}$ & $E^\text{intr}_\text{HFF}$ & $E^\text{intr}_\text{LFF}$ & $\Delta E$ \\ \hline 
0 & -1799.74 & 92.00 & 146.87 & 0.88 & 0.56 & 0.53 & 1.64 & 137.94 & 13.66 & 163.69 & 2036 & 52.85 & 39.15 & 85.02 & 61.85 & -1133.12 & -830.35 & -0.03\\ \hline
4 & -1799.77 & 92.00 & 146.90 & 1.00 & 0.55 & -0.05 & -0.05 & 137.90 & 13.67 & 164.58 & 3070 & 52.22 & 39.78 & 83.83 & 63.07 & -1121.63 & -842.74 & -0.02\\ \hline
7233 & -1793.11 & 92.00 & 146.88 & 7.53 & 0.56 & 1.15 & 1.15 & 140.12 & 13.59 & 166.54 & 2029 & 51.99 & 40.01 & 83.71 & 63.17 & -1115.50 & -844.17 & -0.03\\ \hline
12628 & -1791.09 & 92.00 & 146.89 & 9.58 & 0.55 & 0.18 & 0.76 & 139.32 & 13.53 & 168.23 & 1978 & 52.01 & 39.99 & 83.36 & 63.53 & -1114.72 & -844.62 & -0.02\\ \hline
EE & -1801.26 & 92.00 & 147.00 & - & - & - & - & 138.00 & 13.68 & 170.80 & 2004 & 51.66 & 40.34 & 83.24 & 63.76 & -1115.67 & -856.42 & -0.03\\ \hline
\end{tabular}
\caption{$^{239}$U Run 13}
%\end{table}
%\begin{table}[h]
\centering
\begin{tabular}{|c|c|c|c|c|c|c|c|c|c|c|c|c|c|c|c|c|c|c|}
\hline
Flip & $E$ & $Z$ & $N$ & $E_\text{QP}$ & $n_k$ & $\langle j_z \rangle$ & $j_z$ & $Q_{20}$ & $Q_{30}$ & TKE & $t_\text{scission}$ & $Z_\text{HFF}$ & $Z_\text{LFF}$ & $N_\text{HFF}$ & $N_\text{LFF}$ & $E^\text{intr}_\text{HFF}$ & $E^\text{intr}_\text{LFF}$ & $\Delta E$ \\ \hline 
0 & -1800.09 & 92.00 & 146.91 & 0.84 & 0.55 & 0.68 & 1.78 & 142.58 & 15.23 & 167.74 & 2172 & 52.07 & 39.93 & 84.08 & 62.82 & -1122.05 & -845.82 & -0.04\\ \hline
2 & -1801.52 & 92.00 & 147.20 & 0.87 & 0.40 & -1.69 & -3.19 & 142.78 & 15.25 & 168.80 & 2294 & 51.93 & 40.07 & 83.98 & 63.22 & -1120.32 & -850.03 & -0.03\\ \hline
4 & -1800.69 & 92.00 & 147.04 & 0.93 & 0.48 & -0.79 & -1.86 & 142.71 & 15.24 & 175.97 & 2527 & 51.27 & 40.73 & 83.13 & 63.92 & -1111.56 & -865.15 & -0.05\\ \hline
6 & -1799.06 & 92.00 & 146.73 & 0.99 & 0.63 & 0.20 & 0.20 & 142.43 & 15.23 & 172.92 & 2340 & 51.70 & 40.30 & 82.95 & 63.78 & -1114.90 & -857.11 & -0.03\\ \hline
EE & -1801.39 & 92.00 & 147.00 & - & - & - & - & 142.67 & 15.24 & 167.79 & 2023 & 51.82 & 40.18 & 83.73 & 63.27 & -1118.28 & -850.94 & -0.03\\ \hline
\end{tabular}
\caption{$^{239}$U Run 14}
%\end{table}

\end{sidewaystable}

\clearpage
\begin{sidewaystable}[h]
\vspace{3.5in}
%\begin{table}[h]
\centering
\begin{tabular}{|c|c|c|c|c|c|c|c|c|c|c|c|c|c|c|c|c|c|c|c|c|c|c|c|}
\hline
Flip N & Flip P & $E$ & $Z$ & $N$ & $E_\text{QP,N}$ & $E_\text{QP,P}$ & $n_{k,N}$ & $n_{k,P}$ & $\langle j_z \rangle_N$ & $\langle j_z \rangle_P$ & $j_{z,N}$ & $j_{z,P}$ & $Q_{20}$ & $Q_{30}$ & TKE & $t_\text{scission}$ & $Z_\text{HFF}$ & $Z_\text{LFF}$ & $N_\text{HFF}$ & $N_\text{LFF}$ & $E^\text{intr}_\text{HFF}$ & $E^\text{intr}_\text{LFF}$ & $\Delta E$ \\ \hline 
7681 & 7680 & -1799.40 & 93.55 & 144.94 & 0.97 & 0.67 & 0.52 & 0.24 & 0.00 & -2.53 & 0.00 & -2.54 & 186.52 & 8.49 & 184.02 & 3406 & 51.52 & 42.03 & 81.91 & 63.03 & -1106.53 & -876.96 & -0.07\\ \hline
7681 & 7682 & -1791.68 & 92.35 & 144.94 & 0.97 & 0.83 & 0.52 & 0.84 & 0.00 & 1.34 & 0.00 & 1.37 & 185.53 & 8.53 & 168.19 & 1823 & 50.98 & 41.37 & 81.20 & 63.74 & -1087.68 & -872.23 & -0.04\\ \hline
7686 & 7680 & -1800.14 & 93.55 & 145.07 & 1.01 & 0.67 & 0.45 & 0.24 & -0.37 & -2.53 & -0.37 & -2.54 & 186.53 & 8.53 & 171.72 & 5332 & 52.77 & 40.77 & 82.49 & 62.58 & -1118.11 & -853.82 & -0.07\\ \hline
7686 & 7682 & -1792.40 & 92.35 & 145.07 & 1.01 & 0.83 & 0.45 & 0.84 & -0.37 & 1.34 & -0.37 & 1.37 & 185.54 & 8.57 & 168.63 & 1661 & 51.40 & 40.95 & 81.86 & 63.20 & -1098.56 & -862.52 & -0.05\\ \hline
EE & EE & -1797.99 & 93.03 & 144.97 & - & - & - & - & - & - & - & - & 186.18 & 8.53 & 172.73 & 1506 & 51.64 & 41.39 & 81.38 & 63.59 & -1099.58 & -871.19 & -0.05\\ \hline
\end{tabular}
\caption{$^{238}$Np Run 00 Asymmetric}
%\end{table}
%\begin{table}[h]
\centering
\begin{tabular}{|c|c|c|c|c|c|c|c|c|c|c|c|c|c|c|c|c|c|c|c|c|c|c|c|}
\hline
Flip N & Flip P & $E$ & $Z$ & $N$ & $E_\text{QP,N}$ & $E_\text{QP,P}$ & $n_{k,N}$ & $n_{k,P}$ & $\langle j_z \rangle_N$ & $\langle j_z \rangle_P$ & $j_{z,N}$ & $j_{z,P}$ & $Q_{20}$ & $Q_{30}$ & TKE & $t_\text{scission}$ & $Z_\text{HFF}$ & $Z_\text{LFF}$ & $N_\text{HFF}$ & $N_\text{LFF}$ & $E^\text{intr}_\text{HFF}$ & $E^\text{intr}_\text{LFF}$ & $\Delta E$ \\ \hline 
7681 & 7683 & -1794.04 & 92.96 & 144.69 & 0.67 & 0.54 & 0.63 & 0.52 & -1.06 & 1.82 & -1.23 & 1.92 & 185.33 & 26.98 & 181.13 & 1351 & 52.13 & 40.83 & 82.28 & 62.41 & -1115.50 & -859.73 & -0.06\\ \hline
7681 & 9073 & -1784.59 & 92.85 & 144.69 & 0.67 & 9.26 & 0.63 & 0.58 & -1.06 & -1.36 & -1.23 & -1.36 & 185.39 & 26.95 & 175.84 & 1254 & 52.29 & 40.56 & 82.11 & 62.58 & -1111.70 & -848.80 & -0.06\\ \hline
16134 & 7683 & -1782.47 & 92.96 & 144.83 & 12.93 & 0.54 & 0.56 & 0.52 & -0.21 & 1.82 & -0.27 & 1.92 & 186.40 & 27.00 & 172.75 & 1454 & 52.28 & 40.68 & 82.26 & 62.57 & -1108.29 & -846.97 & -0.03\\ \hline
16134 & 9073 & -1773.01 & 92.85 & 144.83 & 12.93 & 9.26 & 0.56 & 0.58 & -0.21 & -1.36 & -0.27 & -1.36 & 186.46 & 26.96 & 174.19 & 1319 & 52.16 & 40.69 & 82.29 & 62.54 & -1106.45 & -840.79 & -0.04\\ \hline
EE & EE & -1796.88 & 93.00 & 144.94 & - & - & - & - & - & - & - & - & 185.61 & 26.96 & 175.94 & 1331 & 52.03 & 40.97 & 82.23 & 62.72 & -1110.90 & -861.97 & -0.05\\ \hline
\end{tabular}
\caption{$^{238}$Np Run 01 Asymmetric}
%\end{table}
%\begin{table}[h]
\centering
\begin{tabular}{|c|c|c|c|c|c|c|c|c|c|c|c|c|c|c|c|c|c|c|c|c|c|c|c|}
\hline
Flip N & Flip P & $E$ & $Z$ & $N$ & $E_\text{QP,N}$ & $E_\text{QP,P}$ & $n_{k,N}$ & $n_{k,P}$ & $\langle j_z \rangle_N$ & $\langle j_z \rangle_P$ & $j_{z,N}$ & $j_{z,P}$ & $Q_{20}$ & $Q_{30}$ & TKE & $t_\text{scission}$ & $Z_\text{HFF}$ & $Z_\text{LFF}$ & $N_\text{HFF}$ & $N_\text{LFF}$ & $E^\text{intr}_\text{HFF}$ & $E^\text{intr}_\text{LFF}$ & $\Delta E$ \\ \hline 
7682 & 7682 & -1795.41 & 93.01 & 144.89 & 0.99 & 0.76 & 0.55 & 0.50 & 0.68 & -0.29 & 0.80 & -0.30 & 159.79 & 8.77 & 175.87 & 1945 & 51.86 & 41.15 & 82.03 & 62.86 & -1107.36 & -864.00 & -0.07\\ \hline
7682 & 8384 & -1790.24 & 92.90 & 144.89 & 0.99 & 5.19 & 0.55 & 0.55 & 0.68 & -1.06 & 0.80 & -1.94 & 159.77 & 8.78 & 175.61 & 1971 & 51.68 & 41.23 & 81.35 & 63.54 & -1097.64 & -868.27 & -0.07\\ \hline
12282 & 7682 & -1787.55 & 93.01 & 145.05 & 9.74 & 0.76 & 0.47 & 0.50 & 0.17 & -0.29 & 0.96 & -0.30 & 161.82 & 8.96 & 180.01 & 1823 & 52.07 & 40.94 & 82.51 & 62.54 & -1110.82 & -856.78 & -0.05\\ \hline
12282 & 8384 & -1782.39 & 92.90 & 145.05 & 9.74 & 5.19 & 0.47 & 0.55 & 0.17 & -1.06 & 0.96 & -1.94 & 161.79 & 8.97 & 178.82 & 2185 & 51.78 & 41.12 & 82.60 & 62.45 & -1104.12 & -857.15 & -0.06\\ \hline
EE & EE & -1797.71 & 93.01 & 145.00 & - & - & - & - & - & - & - & - & 159.89 & 8.77 & 173.76 & 2715 & 52.15 & 40.86 & 82.69 & 62.31 & -1114.35 & -857.17 & -0.06\\ \hline
\end{tabular}
\caption{$^{238}$Np Run 02 Asymmetric}
%\end{table}
%\begin{table}[h]
\centering
\begin{tabular}{|c|c|c|c|c|c|c|c|c|c|c|c|c|c|c|c|c|c|c|c|c|c|c|c|}
\hline
Flip N & Flip P & $E$ & $Z$ & $N$ & $E_\text{QP,N}$ & $E_\text{QP,P}$ & $n_{k,N}$ & $n_{k,P}$ & $\langle j_z \rangle_N$ & $\langle j_z \rangle_P$ & $j_{z,N}$ & $j_{z,P}$ & $Q_{20}$ & $Q_{30}$ & TKE & $t_\text{scission}$ & $Z_\text{HFF}$ & $Z_\text{LFF}$ & $N_\text{HFF}$ & $N_\text{LFF}$ & $E^\text{intr}_\text{HFF}$ & $E^\text{intr}_\text{LFF}$ & $\Delta E$ \\ \hline 
7680 & 7680 & -1796.14 & 92.99 & 145.15 & 0.71 & 0.61 & 0.44 & 0.50 & -0.68 & -0.95 & -1.64 & -2.12 & 134.81 & 12.64 & 174.79 & 1971 & 52.80 & 40.20 & 83.16 & 61.99 & -1124.94 & -846.05 & -0.06\\ \hline
7680 & 7684 & -1795.19 & 92.84 & 145.15 & 0.71 & 0.63 & 0.44 & 0.58 & -0.68 & -0.63 & -1.64 & -0.72 & 134.78 & 12.64 & 164.62 & 2107 & 53.47 & 39.37 & 84.96 & 60.20 & -1137.93 & -821.92 & -0.04\\ \hline
9055 & 7680 & -1790.31 & 92.99 & 144.97 & 5.48 & 0.61 & 0.54 & 0.50 & 0.17 & -0.95 & 0.32 & -2.12 & 134.85 & 12.70 & 177.41 & 1881 & 52.64 & 40.36 & 82.89 & 62.08 & -1120.87 & -846.92 & -0.06\\ \hline
9055 & 7684 & -1789.36 & 92.84 & 144.97 & 5.48 & 0.63 & 0.54 & 0.58 & 0.17 & -0.63 & 0.32 & -0.72 & 134.82 & 12.70 & 172.35 & 2139 & 52.98 & 39.85 & 84.30 & 60.67 & -1130.18 & -831.57 & -0.03\\ \hline
EE & EE & -1796.84 & 93.00 & 145.04 & - & - & - & - & - & - & - & - & 134.81 & 12.63 & 179.78 & 1965 & 52.17 & 40.83 & 82.33 & 62.71 & -1115.79 & -860.90 & -0.06\\ \hline
\end{tabular}
\caption{$^{238}$Np Run 03 Asymmetric}
%\end{table}
%\begin{table}[h]
\centering
\begin{tabular}{|c|c|c|c|c|c|c|c|c|c|c|c|c|c|c|c|c|c|c|c|c|c|c|c|}
\hline
Flip N & Flip P & $E$ & $Z$ & $N$ & $E_\text{QP,N}$ & $E_\text{QP,P}$ & $n_{k,N}$ & $n_{k,P}$ & $\langle j_z \rangle_N$ & $\langle j_z \rangle_P$ & $j_{z,N}$ & $j_{z,P}$ & $Q_{20}$ & $Q_{30}$ & TKE & $t_\text{scission}$ & $Z_\text{HFF}$ & $Z_\text{LFF}$ & $N_\text{HFF}$ & $N_\text{LFF}$ & $E^\text{intr}_\text{HFF}$ & $E^\text{intr}_\text{LFF}$ & $\Delta E$ \\ \hline 
7680 & 7683 & -1797.74 & 92.97 & 145.48 & 0.75 & 0.67 & 0.30 & 0.51 & -1.73 & 0.34 & -1.83 & 0.49 & 135.19 & 9.77 & 161.99 & 2055 & 53.84 & 39.14 & 85.75 & 59.73 & -1143.36 & -816.40 & -0.03\\ \hline
7680 & 9730 & -1786.99 & 92.93 & 145.48 & 0.75 & 11.10 & 0.30 & 0.53 & -1.73 & -0.57 & -1.83 & -0.72 & 134.91 & 9.76 & 175.78 & 1978 & 52.19 & 40.74 & 83.24 & 62.24 & -1110.28 & -852.56 & -0.07\\ \hline
7680 & 9732 & -1787.72 & 93.05 & 145.48 & 0.75 & 11.13 & 0.30 & 0.47 & -1.73 & 0.67 & -1.83 & 0.88 & 134.98 & 9.77 & 172.60 & 2023 & 52.67 & 40.38 & 83.50 & 61.98 & -1115.10 & -845.28 & -0.06\\ \hline
7683 & 7683 & -1796.53 & 92.97 & 145.27 & 0.82 & 0.67 & 0.40 & 0.51 & 0.06 & 0.34 & 0.06 & 0.49 & 135.26 & 9.77 & 170.61 & 2585 & 52.40 & 40.57 & 83.29 & 61.98 & -1120.27 & -846.89 & -0.03\\ \hline
7683 & 9730 & -1785.76 & 92.93 & 145.27 & 0.82 & 11.10 & 0.40 & 0.53 & 0.06 & -0.57 & 0.06 & -0.72 & 134.99 & 9.77 & 174.67 & 1868 & 52.18 & 40.75 & 82.73 & 62.54 & -1110.54 & -849.96 & -0.06\\ \hline
7683 & 9732 & -1786.47 & 93.05 & 145.27 & 0.82 & 11.13 & 0.40 & 0.47 & 0.06 & 0.67 & 0.06 & 0.88 & 135.05 & 9.78 & 172.16 & 1887 & 52.55 & 40.50 & 83.14 & 62.13 & -1116.23 & -842.46 & -0.06\\ \hline
EE & EE & -1797.04 & 92.99 & 145.07 & - & - & - & - & - & - & - & - & 135.09 & 9.75 & 178.38 & 1965 & 52.16 & 40.83 & 82.19 & 62.89 & -1114.78 & -860.69 & -0.05\\ \hline
\end{tabular}
\caption{$^{238}$Np Run 04 Asymmetric}
%\end{table}
\end{sidewaystable}

\clearpage
\begin{sidewaystable}[h]
\vspace{3.5in}
%\begin{table}[h]
\centering
\begin{tabular}{|c|c|c|c|c|c|c|c|c|c|c|c|c|c|c|c|c|c|c|c|c|c|c|c|}
\hline
Flip N & Flip P & $E$ & $Z$ & $N$ & $E_\text{QP,N}$ & $E_\text{QP,P}$ & $n_{k,N}$ & $n_{k,P}$ & $\langle j_z \rangle_N$ & $\langle j_z \rangle_P$ & $j_{z,N}$ & $j_{z,P}$ & $Q_{20}$ & $Q_{30}$ & TKE & $t_\text{scission}$ & $Z_\text{HFF}$ & $Z_\text{LFF}$ & $N_\text{HFF}$ & $N_\text{LFF}$ & $E^\text{intr}_\text{HFF}$ & $E^\text{intr}_\text{LFF}$ & $\Delta E$ \\ \hline 
7681 & 7680 & -1796.74 & 93.20 & 144.88 & 0.65 & 0.65 & 0.55 & 0.42 & 0.77 & -1.86 & 0.79 & -1.94 & 202.00 & 31.67 & 177.59 & 1325 & 52.28 & 40.93 & 82.11 & 62.77 & -1116.02 & -858.36 & -0.06\\ \hline
7681 & 7682 & -1796.10 & 93.10 & 144.88 & 0.65 & 0.68 & 0.55 & 0.47 & 0.77 & 0.05 & 0.79 & 0.29 & 201.93 & 31.67 & 176.84 & 1493 & 52.72 & 40.38 & 82.42 & 62.46 & -1121.33 & -851.67 & -0.06\\ \hline
12296 & 7680 & -1787.54 & 93.20 & 144.85 & 9.60 & 0.65 & 0.56 & 0.42 & -2.91 & -1.86 & -3.47 & -1.94 & 203.45 & 31.42 & 174.56 & 1254 & 52.11 & 41.09 & 82.04 & 62.80 & -1104.99 & -857.14 & -0.04\\ \hline
12296 & 7682 & -1786.89 & 93.10 & 144.85 & 9.60 & 0.68 & 0.56 & 0.47 & -2.91 & 0.05 & -3.47 & 0.29 & 203.38 & 31.42 & 174.79 & 1357 & 52.65 & 40.46 & 82.33 & 62.51 & -1111.49 & -850.23 & -0.04\\ \hline
EE & EE & -1797.52 & 93.04 & 144.97 & - & - & - & - & - & - & - & - & 201.95 & 31.65 & 175.57 & 1280 & 52.34 & 40.70 & 82.44 & 62.54 & -1117.57 & -855.58 & -0.06\\ \hline
\end{tabular}
\caption{$^{238}$Np Run 05 Asymmetric}
%\end{table}
%\begin{table}[h]
\centering
\begin{tabular}{|c|c|c|c|c|c|c|c|c|c|c|c|c|c|c|c|c|c|c|c|c|c|c|c|}
\hline
Flip N & Flip P & $E$ & $Z$ & $N$ & $E_\text{QP,N}$ & $E_\text{QP,P}$ & $n_{k,N}$ & $n_{k,P}$ & $\langle j_z \rangle_N$ & $\langle j_z \rangle_P$ & $j_{z,N}$ & $j_{z,P}$ & $Q_{20}$ & $Q_{30}$ & TKE & $t_\text{scission}$ & $Z_\text{HFF}$ & $Z_\text{LFF}$ & $N_\text{HFF}$ & $N_\text{LFF}$ & $E^\text{intr}_\text{HFF}$ & $E^\text{intr}_\text{LFF}$ & $\Delta E$ \\ \hline 
7686 & 7681 & -1797.28 & 93.17 & 145.02 & 0.97 & 0.66 & 0.50 & 0.42 & -1.19 & -1.00 & -1.83 & -1.02 & 151.30 & 17.36 & 180.54 & 1855 & 52.24 & 40.92 & 82.75 & 62.27 & -1119.18 & -858.70 & -0.06\\ \hline
7686 & 9079 & -1786.11 & 92.88 & 145.02 & 0.97 & 10.09 & 0.50 & 0.57 & -1.19 & -0.02 & -1.83 & -1.10 & 151.14 & 17.37 & 180.55 & 1713 & 51.81 & 41.07 & 82.69 & 62.33 & -1109.57 & -857.16 & -0.07\\ \hline
8399 & 7681 & -1793.31 & 93.17 & 145.01 & 4.88 & 0.66 & 0.50 & 0.42 & -0.87 & -1.00 & -1.32 & -1.02 & 151.48 & 17.39 & 176.59 & 2081 & 52.46 & 40.70 & 83.01 & 62.01 & -1117.01 & -852.96 & -0.06\\ \hline
8399 & 9079 & -1782.14 & 92.88 & 145.01 & 4.88 & 10.09 & 0.50 & 0.57 & -0.87 & -0.02 & -1.32 & -1.10 & 151.32 & 17.40 & 175.64 & 1887 & 52.37 & 40.51 & 83.51 & 61.51 & -1112.95 & -844.89 & -0.05\\ \hline
9042 & 7681 & -1793.24 & 93.17 & 145.02 & 5.00 & 0.66 & 0.50 & 0.42 & 0.83 & -1.00 & 1.41 & -1.02 & 151.46 & 17.38 & 173.96 & 1861 & 53.20 & 39.97 & 84.11 & 60.91 & -1132.28 & -834.96 & -0.04\\ \hline
9042 & 9079 & -1782.08 & 92.88 & 145.02 & 5.00 & 10.09 & 0.50 & 0.57 & 0.83 & -0.02 & 1.41 & -1.10 & 151.29 & 17.39 & 175.05 & 1752 & 52.26 & 40.62 & 83.36 & 61.66 & -1111.97 & -845.22 & -0.06\\ \hline
EE & EE & -1798.01 & 93.01 & 145.02 & - & - & - & - & - & - & - & - & 151.22 & 17.34 & 174.39 & 1797 & 52.73 & 40.28 & 83.42 & 61.60 & -1125.03 & -847.42 & -0.05\\ \hline
\end{tabular}
\caption{$^{238}$Np Run 06 Asymmetric}
%\end{table}
%\begin{table}[h]
\centering
\begin{tabular}{|c|c|c|c|c|c|c|c|c|c|c|c|c|c|c|c|c|c|c|c|c|c|c|c|}
\hline
Flip N & Flip P & $E$ & $Z$ & $N$ & $E_\text{QP,N}$ & $E_\text{QP,P}$ & $n_{k,N}$ & $n_{k,P}$ & $\langle j_z \rangle_N$ & $\langle j_z \rangle_P$ & $j_{z,N}$ & $j_{z,P}$ & $Q_{20}$ & $Q_{30}$ & TKE & $t_\text{scission}$ & $Z_\text{HFF}$ & $Z_\text{LFF}$ & $N_\text{HFF}$ & $N_\text{LFF}$ & $E^\text{intr}_\text{HFF}$ & $E^\text{intr}_\text{LFF}$ & $\Delta E$ \\ \hline 
7680 & 7683 & -1795.60 & 93.06 & 144.92 & 0.92 & 0.80 & 0.55 & 0.47 & 4.49 & 0.68 & 4.50 & 2.87 & 149.23 & 8.63 & 176.48 & 1764 & 52.15 & 40.91 & 81.67 & 63.25 & -1111.13 & -861.00 & -0.06\\ \hline
7680 & 7687 & -1795.87 & 93.11 & 144.92 & 0.92 & 0.84 & 0.55 & 0.45 & 4.49 & -0.70 & 4.50 & -0.72 & 149.28 & 8.65 & 177.21 & 1577 & 52.02 & 41.08 & 81.79 & 63.14 & -1110.71 & -862.41 & -0.05\\ \hline
7684 & 7683 & -1796.16 & 93.06 & 145.03 & 0.97 & 0.80 & 0.49 & 0.47 & 0.73 & 0.68 & 2.01 & 2.87 & 149.27 & 8.67 & 169.47 & 2728 & 53.83 & 39.23 & 85.07 & 59.96 & -1141.59 & -824.08 & -0.04\\ \hline
7684 & 7687 & -1796.43 & 93.11 & 145.03 & 0.97 & 0.84 & 0.49 & 0.45 & 0.73 & -0.70 & 2.01 & -0.72 & 149.32 & 8.68 & 168.43 & 1849 & 53.05 & 40.05 & 83.60 & 61.44 & -1126.51 & -838.39 & -0.04\\ \hline
10375 & 7683 & -1789.10 & 93.06 & 144.96 & 7.55 & 0.80 & 0.53 & 0.47 & -0.47 & 0.68 & -0.48 & 2.87 & 152.33 & 8.02 & 183.13 & 1726 & 52.01 & 41.05 & 82.52 & 62.44 & -1113.78 & -858.51 & -0.05\\ \hline
10375 & 7687 & -1789.38 & 93.11 & 144.96 & 7.55 & 0.84 & 0.53 & 0.45 & -0.47 & -0.70 & -0.48 & -0.72 & 152.38 & 8.03 & 178.40 & 1564 & 52.13 & 40.98 & 83.32 & 61.64 & -1116.36 & -851.45 & -0.04\\ \hline
11658 & 7683 & -1787.88 & 93.06 & 145.06 & 9.38 & 0.80 & 0.48 & 0.47 & -0.46 & 0.68 & -0.99 & 2.87 & 151.00 & 8.53 & 179.31 & 1874 & 52.46 & 40.60 & 82.67 & 62.40 & -1114.82 & -852.41 & -0.04\\
\hline
\end{tabular}
\caption{$^{238}$Np Run 07 Asymmetric}
%\end{table}
%\begin{table}[h]
\centering
\begin{tabular}{|c|c|c|c|c|c|c|c|c|c|c|c|c|c|c|c|c|c|c|c|c|c|c|c|}
\hline
Flip N & Flip P & $E$ & $Z$ & $N$ & $E_\text{QP,N}$ & $E_\text{QP,P}$ & $n_{k,N}$ & $n_{k,P}$ & $\langle j_z \rangle_N$ & $\langle j_z \rangle_P$ & $j_{z,N}$ & $j_{z,P}$ & $Q_{20}$ & $Q_{30}$ & TKE & $t_\text{scission}$ & $Z_\text{HFF}$ & $Z_\text{LFF}$ & $N_\text{HFF}$ & $N_\text{LFF}$ & $E^\text{intr}_\text{HFF}$ & $E^\text{intr}_\text{LFF}$ & $\Delta E$ \\ \hline 
7680 & 7685 & -1793.40 & 92.71 & 144.84 & 1.01 & 0.75 & 0.57 & 0.66 & 1.32 & 1.21 & 2.01 & 1.23 & 177.33 & 8.76 & 172.91 & 1752 & 52.04 & 40.66 & 82.71 & 62.13 & -1111.54 & -854.84 & -0.06\\ \hline
7683 & 7685 & -1793.85 & 92.71 & 144.92 & 1.03 & 0.75 & 0.53 & 0.66 & 2.17 & 1.21 & 2.39 & 1.23 & 177.44 & 8.73 & 172.50 & 1668 & 51.97 & 40.74 & 82.07 & 62.84 & -1108.56 & -857.85 & -0.06\\ \hline
EE & EE & -1797.92 & 93.02 & 144.98 & - & - & - & - & - & - & - & - & 177.77 & 8.76 & 171.30 & 1668 & 52.28 & 40.74 & 82.03 & 62.95 & -1107.67 & -861.61 & -0.06\\ \hline
\end{tabular}
\caption{$^{238}$Np Run 08 Asymmetric}
%\end{table}
%\begin{table}[h]
\centering
\begin{tabular}{|c|c|c|c|c|c|c|c|c|c|c|c|c|c|c|c|c|c|c|c|c|c|c|c|}
\hline
Flip N & Flip P & $E$ & $Z$ & $N$ & $E_\text{QP,N}$ & $E_\text{QP,P}$ & $n_{k,N}$ & $n_{k,P}$ & $\langle j_z \rangle_N$ & $\langle j_z \rangle_P$ & $j_{z,N}$ & $j_{z,P}$ & $Q_{20}$ & $Q_{30}$ & TKE & $t_\text{scission}$ & $Z_\text{HFF}$ & $Z_\text{LFF}$ & $N_\text{HFF}$ & $N_\text{LFF}$ & $E^\text{intr}_\text{HFF}$ & $E^\text{intr}_\text{LFF}$ & $\Delta E$ \\ \hline 
9726 & 7680 & -1792.47 & 93.22 & 145.10 & 6.55 & 0.44 & 0.46 & 0.40 & 0.55 & 1.56 & 0.94 & 1.64 & 171.62 & 22.21 & 174.09 & 1467 & 52.24 & 40.98 & 82.07 & 63.02 & -1106.80 & -859.79 & -0.04\\ \hline
9726 & 9040 & -1784.01 & 92.86 & 145.10 & 6.55 & 6.67 & 0.46 & 0.58 & 0.55 & -0.09 & 0.94 & -0.10 & 171.37 & 22.20 & 174.45 & 1370 & 51.92 & 40.93 & 82.09 & 63.01 & -1100.98 & -857.51 & -0.03\\ \hline
9728 & 7680 & -1791.90 & 93.22 & 145.00 & 6.57 & 0.44 & 0.50 & 0.40 & -0.55 & 1.56 & -0.95 & 1.64 & 170.61 & 22.14 & 174.10 & 1467 & 52.28 & 40.94 & 82.06 & 62.95 & -1107.34 & -858.70 & -0.03\\ \hline
9728 & 9040 & -1783.45 & 92.86 & 145.00 & 6.57 & 6.67 & 0.50 & 0.58 & -0.55 & -0.09 & -0.95 & -0.10 & 170.36 & 22.12 & 174.33 & 1370 & 51.95 & 40.91 & 82.07 & 62.93 & -1101.16 & -856.65 & -0.03\\ \hline
EE & EE & -1797.80 & 93.02 & 145.01 & - & - & - & - & - & - & - & - & 168.60 & 22.28 & 172.80 & 1461 & 52.30 & 40.73 & 82.32 & 62.69 & -1112.19 & -858.45 & -0.04\\ \hline
\end{tabular}
\caption{$^{238}$Np Run 09 Asymmetric}
%\end{table}
\end{sidewaystable}

\clearpage
\begin{sidewaystable}[h]
\vspace{3.5in}
%\begin{table}[h]
\centering
\begin{tabular}{|c|c|c|c|c|c|c|c|c|c|c|c|c|c|c|c|c|c|c|c|c|c|c|c|}
\hline
Flip N & Flip P & $E$ & $Z$ & $N$ & $E_\text{QP,N}$ & $E_\text{QP,P}$ & $n_{k,N}$ & $n_{k,P}$ & $\langle j_z \rangle_N$ & $\langle j_z \rangle_P$ & $j_{z,N}$ & $j_{z,P}$ & $Q_{20}$ & $Q_{30}$ & TKE & $t_\text{scission}$ & $Z_\text{HFF}$ & $Z_\text{LFF}$ & $N_\text{HFF}$ & $N_\text{LFF}$ & $E^\text{intr}_\text{HFF}$ & $E^\text{intr}_\text{LFF}$ & $\Delta E$ \\ \hline 
36 & 7682 & -1779.23 & 93.03 & 145.06 & 19.33 & 0.89 & 0.48 & 0.49 & -0.80 & -0.72 & -1.35 & -2.84 & 206.38 & 0.23 & 152.34 & 5061 & 49.58 & 43.45 & 77.84 & 67.22 & -1031.25 & -900.32 & 0.01\\ \hline
38 & 7682 & -1778.86 & 93.03 & 145.00 & 19.35 & 0.89 & 0.51 & 0.49 & 0.86 & -0.72 & 1.35 & -2.84 & 206.31 & 0.20 & 149.74 & 4763 & 49.48 & 43.55 & 77.83 & 67.17 & -1028.20 & -900.38 & 0.01\\ \hline
EE & EE & -1799.07 & 93.01 & 145.02 & - & - & - & - & - & - & - & - & 204.63 & 0.33 & 148.24 & 5772 & 49.90 & 43.10 & 78.46 & 66.56 & -1043.28 & -904.04 & -0.02\\ \hline
\end{tabular}
\caption{$^{238}$Np Run 01 Symmetric}
%\end{table}

%\begin{table}[h]
\centering
\begin{tabular}{|c|c|c|c|c|c|c|c|c|c|c|c|c|c|c|c|c|c|c|}
\hline
Flip & $E$ & $Z$ & $N$ & $E_\text{QP}$ & $n_k$ & $\langle j_z \rangle$ & $j_z$ & $Q_{20}$ & $Q_{30}$ & TKE & $t_\text{scission}$ & $Z_\text{HFF}$ & $Z_\text{LFF}$ & $N_\text{HFF}$ & $N_\text{LFF}$ & $E^\text{intr}_\text{HFF}$ & $E^\text{intr}_\text{LFF}$ & $\Delta E$ \\ \hline 
0 & -1823.91 & 94.00 & 148.81 & 0.97 & 0.59 & -4.49 & -4.50 & 158.35 & 7.85 & 172.87 & 1965 & 52.20 & 41.80 & 83.97 & 64.85 & -1116.85 & -879.94 & -0.01\\ \hline
4 & -1825.78 & 94.00 & 149.17 & 1.06 & 0.41 & -0.41 & -1.64 & 158.49 & 7.83 & 170.12 & 1603 & 52.56 & 41.44 & 84.32 & 64.86 & -1126.91 & -869.02 & -0.02\\ \hline
8 & -1824.23 & 94.00 & 148.90 & 1.11 & 0.55 & -2.30 & -2.54 & 158.33 & 7.90 & 165.75 & 2210 & 53.61 & 40.39 & 85.87 & 63.03 & -1140.84 & -849.15 & 0.00\\ \hline
EE & -1825.86 & 94.00 & 149.00 & –– & –– & –– & –– & 158.41 & 7.92 & 168.75 & 1661 & 53.24 & 40.76 & 84.98 & 64.02 & -1135.29 & -859.33 & -0.00\\ \hline
\end{tabular}
\caption{$^{243}$Pu Run 00 Asymmetric}
%\end{table}

%\begin{table}[h]
\centering
\begin{tabular}{|c|c|c|c|c|c|c|c|c|c|c|c|c|c|c|c|c|c|c|}
\hline
Flip & $E$ & $Z$ & $N$ & $E_\text{QP}$ & $n_k$ & $\langle j_z \rangle$ & $j_z$ & $Q_{20}$ & $Q_{30}$ & TKE & $t_\text{scission}$ & $Z_\text{HFF}$ & $Z_\text{LFF}$ & $N_\text{HFF}$ & $N_\text{LFF}$ & $E^\text{intr}_\text{HFF}$ & $E^\text{intr}_\text{LFF}$ & $\Delta E$ \\ \hline 
0 & -1823.72 & 94.00 & 148.91 & 0.94 & 0.54 & 1.32 & 1.56 & 139.73 & 10.19 & 169.70 & 2618 & 52.79 & 41.21 & 84.39 & 64.52 & -1128.50 & -864.95 & -0.03\\ \hline
2 & -1824.28 & 94.00 & 149.03 & 1.02 & 0.48 & -1.79 & -2.89 & 139.79 & 10.20 & 174.06 & 4033 & 52.35 & 41.65 & 83.68 & 65.36 & -1120.05 & -878.31 & -0.02\\ \hline
8 & -1823.59 & 94.00 & 148.92 & 1.12 & 0.54 & -1.41 & -1.89 & 139.70 & 10.20 & 178.34 & 2773 & 52.47 & 41.53 & 83.82 & 65.10 & -1122.64 & -879.32 & -0.03\\ \hline
EE & -1825.12 & 94.00 & 149.00 & –– & –– & –– & –– & 139.76 & 10.20 & 171.26 & 3128 & 52.36 & 41.64 & 83.86 & 65.14 & -1120.15 & -876.24 & -0.01\\ \hline
\end{tabular}
\caption{$^{243}$Pu Run 01 Asymmetric}
%\end{table}

%\begin{table}[h]
\centering
\begin{tabular}{|c|c|c|c|c|c|c|c|c|c|c|c|c|c|c|c|c|c|c|}
\hline
Flip & $E$ & $Z$ & $N$ & $E_\text{QP}$ & $n_k$ & $\langle j_z \rangle$ & $j_z$ & $Q_{20}$ & $Q_{30}$ & TKE & $t_\text{scission}$ & $Z_\text{HFF}$ & $Z_\text{LFF}$ & $N_\text{HFF}$ & $N_\text{LFF}$ & $E^\text{intr}_\text{HFF}$ & $E^\text{intr}_\text{LFF}$ & $\Delta E$ \\ \hline 
2 & -1824.57 & 94.00 & 149.03 & 1.01 & 0.48 & -0.72 & -1.74 & 149.61 & 8.11 & 169.69 & 1719 & 52.90 & 41.10 & 84.46 & 64.57 & -1129.87 & -864.40 & -0.02\\ \hline
6 & -1824.80 & 94.00 & 149.09 & 1.09 & 0.45 & 2.37 & 2.67 & 149.68 & 8.11 & 166.50 & 2353 & 53.66 & 40.34 & 85.70 & 63.39 & -1140.57 & -850.73 & -0.00\\ \hline
EE & -1825.40 & 94.00 & 149.00 & –– & –– & –– & –– & 149.64 & 8.16 & 170.85 & 1752 & 52.80 & 41.20 & 84.40 & 64.60 & -1127.58 & -868.68 & -0.01\\ \hline
\end{tabular}
\caption{$^{243}$Pu Run 02 Asymmetric}
%\end{table}
%\begin{table}[h]
\centering
\begin{tabular}{|c|c|c|c|c|c|c|c|c|c|c|c|c|c|c|c|c|c|c|}
\hline
Flip & $E$ & $Z$ & $N$ & $E_\text{QP}$ & $n_k$ & $\langle j_z \rangle$ & $j_z$ & $Q_{20}$ & $Q_{30}$ & TKE & $t_\text{scission}$ & $Z_\text{HFF}$ & $Z_\text{LFF}$ & $N_\text{HFF}$ & $N_\text{LFF}$ & $E^\text{intr}_\text{HFF}$ & $E^\text{intr}_\text{LFF}$ & $\Delta E$ \\ \hline 
0 & -1823.90 & 94.00 & 148.97 & 1.01 & 0.52 & -0.12 & -3.51 & 140.93 & 13.75 & 180.15 & 3167 & 52.04 & 41.96 & 83.52 & 65.45 & -1117.93 & -886.16 & -0.04\\ \hline
4 & -1822.94 & 94.00 & 148.80 & 1.10 & 0.60 & 0.00 & 0.06 & 140.74 & 13.74 & 175.74 & 4408 & 52.40 & 41.60 & 83.67 & 65.14 & -1120.64 & -878.10 & -0.05\\ \hline
6 & -1823.87 & 94.00 & 148.97 & 1.08 & 0.51 & -1.24 & -1.97 & 140.91 & 13.75 & 179.63 & 2799 & 52.24 & 41.76 & 83.48 & 65.49 & -1119.49 & -884.04 & -0.04\\ \hline
EE & -1825.07 & 94.00 & 149.00 & –– & –– & –– & –– & 140.94 & 13.75 & 168.23 & 3742 & 52.74 & 41.26 & 84.77 & 64.23 & -1126.75 & -866.56 & -0.01\\ \hline
\end{tabular}
\caption{$^{243}$Pu Run 03 Asymmetric}
%\end{table}
\end{sidewaystable}

\clearpage
\begin{sidewaystable}[h]
\vspace{3.5in}
%\begin{table}[h]
\centering
\begin{tabular}{|c|c|c|c|c|c|c|c|c|c|c|c|c|c|c|c|c|c|c|}
\hline
Flip & $E$ & $Z$ & $N$ & $E_\text{QP}$ & $n_k$ & $\langle j_z \rangle$ & $j_z$ & $Q_{20}$ & $Q_{30}$ & TKE & $t_\text{scission}$ & $Z_\text{HFF}$ & $Z_\text{LFF}$ & $N_\text{HFF}$ & $N_\text{LFF}$ & $E^\text{intr}_\text{HFF}$ & $E^\text{intr}_\text{LFF}$ & $\Delta E$ \\ \hline 
0 & -1825.64 & 94.00 & 148.99 & 0.88 & 0.50 & -2.84 & -2.96 & 171.23 & 21.89 & 176.16 & 2004 & 52.49 & 41.51 & 83.71 & 65.29 & -1123.74 & -878.09 & -0.03\\ \hline
9032 & -1819.08 & 94.00 & 149.18 & 8.36 & 0.41 & 0.14 & 0.14 & 173.32 & 21.64 & 173.62 & 1687 & 52.60 & 41.40 & 84.15 & 65.03 & -1122.61 & -870.10 & -0.01\\ \hline
10800 & -1816.33 & 94.00 & 148.72 & 8.74 & 0.64 & -2.13 & -2.54 & 172.01 & 22.11 & 173.46 & 1635 & 51.87 & 42.13 & 83.50 & 65.22 & -1110.73 & -879.07 & -0.01\\ \hline
EE & -1826.54 & 94.00 & 149.00 & –– & –– & –– & –– & 171.24 & 21.89 & 174.29 & 1668 & 52.17 & 41.83 & 83.59 & 65.41 & -1118.93 & -881.91 & -0.02\\ \hline
\end{tabular}
\caption{$^{243}$Pu Run 04 Asymmetric}
%\end{table}
%\begin{table}[h]
\centering
\begin{tabular}{|c|c|c|c|c|c|c|c|c|c|c|c|c|c|c|c|c|c|c|}
\hline
Flip & $E$ & $Z$ & $N$ & $E_\text{QP}$ & $n_k$ & $\langle j_z \rangle$ & $j_z$ & $Q_{20}$ & $Q_{30}$ & TKE & $t_\text{scission}$ & $Z_\text{HFF}$ & $Z_\text{LFF}$ & $N_\text{HFF}$ & $N_\text{LFF}$ & $E^\text{intr}_\text{HFF}$ & $E^\text{intr}_\text{LFF}$ & $\Delta E$ \\ \hline 
0 & -1824.29 & 94.00 & 148.78 & 0.96 & 0.61 & 1.42 & 1.45 & 217.24 & 34.72 & 166.54 & 1409 & 53.51 & 40.49 & 85.39 & 63.39 & -1139.01 & -851.85 & -0.03\\ \hline
3630 & -1821.01 & 94.00 & 149.12 & 5.96 & 0.44 & -1.03 & -1.14 & 218.87 & 34.82 & 165.69 & 1254 & 52.72 & 41.28 & 84.15 & 64.97 & -1121.33 & -865.37 & -0.00\\ \hline
3636 & -1820.51 & 94.00 & 149.05 & 6.09 & 0.48 & -0.99 & -1.10 & 220.42 & 34.66 & 165.89 & 1247 & 52.87 & 41.13 & 84.40 & 64.65 & -1124.08 & -862.33 & -0.00\\ \hline
5412 & -1820.11 & 94.00 & 149.02 & 6.31 & 0.49 & 1.12 & 1.21 & 218.43 & 34.20 & 165.43 & 1273 & 53.10 & 40.90 & 84.85 & 64.17 & -1129.10 & -856.45 & -0.00\\ \hline
7236 & -1820.85 & 94.00 & 149.39 & 7.48 & 0.30 & -0.11 & -0.16 & 220.24 & 34.27 & 163.13 & 1286 & 53.49 & 40.51 & 85.66 & 63.73 & -1136.73 & -847.24 & 0.01\\ \hline
\end{tabular}
\caption{$^{243}$Pu Run 05 Asymmetric}
%\end{table}
%\begin{table}[h]
\centering
\begin{tabular}{|c|c|c|c|c|c|c|c|c|c|c|c|c|c|c|c|c|c|c|}
\hline
Flip & $E$ & $Z$ & $N$ & $E_\text{QP}$ & $n_k$ & $\langle j_z \rangle$ & $j_z$ & $Q_{20}$ & $Q_{30}$ & TKE & $t_\text{scission}$ & $Z_\text{HFF}$ & $Z_\text{LFF}$ & $N_\text{HFF}$ & $N_\text{LFF}$ & $E^\text{intr}_\text{HFF}$ & $E^\text{intr}_\text{LFF}$ & $\Delta E$ \\ \hline 
0 & -1824.90 & 94.00 & 148.98 & 0.94 & 0.51 & 2.56 & 3.10 & 154.41 & 17.46 & 181.24 & 2779 & 51.78 & 42.22 & 83.26 & 65.71 & -1114.60 & -891.59 & -0.05\\ \hline
6 & -1824.52 & 94.00 & 148.92 & 1.00 & 0.54 & 1.12 & 1.14 & 154.37 & 17.46 & 171.50 & 4860 & 52.51 & 41.49 & 84.19 & 64.73 & -1122.86 & -873.18 & -0.03\\ \hline
10820 & -1816.73 & 94.00 & 148.96 & 8.97 & 0.52 & -2.80 & -2.94 & 157.27 & 17.03 & 171.02 & 2133 & 52.67 & 41.33 & 84.19 & 64.77 & -1120.92 & -866.85 & -0.02\\ \hline
EE & -1825.93 & 94.00 & 149.00 & –– & –– & –– & –– & 154.42 & 17.46 & 173.39 & 2282 & 52.65 & 41.35 & 84.21 & 64.79 & -1126.21 & -873.16 & -0.03\\ \hline
\end{tabular}
\caption{$^{243}$Pu Run 06 Asymmetric}
%\end{table}
%\begin{table}[h]
\centering
\begin{tabular}{|c|c|c|c|c|c|c|c|c|c|c|c|c|c|c|c|c|c|c|}
\hline
Flip & $E$ & $Z$ & $N$ & $E_\text{QP}$ & $n_k$ & $\langle j_z \rangle$ & $j_z$ & $Q_{20}$ & $Q_{30}$ & TKE & $t_\text{scission}$ & $Z_\text{HFF}$ & $Z_\text{LFF}$ & $N_\text{HFF}$ & $N_\text{LFF}$ & $E^\text{intr}_\text{HFF}$ & $E^\text{intr}_\text{LFF}$ & $\Delta E$ \\ \hline 
7685 & -1824.04 & 94.03 & 148.86 & 0.98 & 0.56 & -2.43 & -2.44 & 181.42 & 6.05 & 168.58 & 2172 & 51.72 & 42.31 & 83.25 & 65.61 & -1105.10 & -887.53 & -0.01\\ \hline
9752 & -1823.98 & 94.03 & 149.97 & 7.18 & 0.00 & 1.12 & 1.14 & 184.41 & 6.76 & 168.82 & 1745 & 51.91 & 42.11 & 83.99 & 65.98 & -1108.94 & -883.86 & 0.00\\ \hline
15505 & -1813.36 & 94.03 & 148.97 & 14.78 & 0.52 & 0.06 & 0.08 & 183.21 & 6.04 & 161.45 & 1732 & 51.49 & 42.53 & 82.66 & 66.31 & -1085.10 & -889.69 & 0.02\\ \hline
EE & -1825.63 & 94.03 & 148.97 & –– & –– & –– & –– & 181.45 & 6.10 & 167.04 & 1700 & 51.69 & 42.34 & 83.02 & 65.96 & -1104.19 & -888.49 & -0.02\\ \hline
\end{tabular}
\caption{$^{243}$Pu Run 07 Asymmetric}
%\end{table}
\end{sidewaystable}

\clearpage
\begin{sidewaystable}[h]
\vspace{3.5in}

%\begin{table}[h]
\centering
\begin{tabular}{|c|c|c|c|c|c|c|c|c|c|c|c|c|c|c|c|c|c|c|}
\hline
Flip & $E$ & $Z$ & $N$ & $E_\text{QP}$ & $n_k$ & $\langle j_z \rangle$ & $j_z$ & $Q_{20}$ & $Q_{30}$ & TKE & $t_\text{scission}$ & $Z_\text{HFF}$ & $Z_\text{LFF}$ & $N_\text{HFF}$ & $N_\text{LFF}$ & $E^\text{intr}_\text{HFF}$ & $E^\text{intr}_\text{LFF}$ & $\Delta E$ \\ \hline 
7681 & -1824.80 & 94.00 & 148.97 & 0.76 & 0.49 & 3.59 & 3.59 & 161.72 & 19.69 & 172.70 & 2353 & 52.61 & 41.39 & 84.22 & 64.74 & -1123.67 & -873.86 & -0.03\\ \hline
7682 & -1824.56 & 94.00 & 148.93 & 0.81 & 0.51 & -2.89 & -3.05 & 161.70 & 19.70 & 177.06 & 2223 & 51.93 & 42.07 & 83.55 & 65.38 & -1115.56 & -886.10 & -0.03\\ \hline
11658 & -1816.50 & 94.00 & 148.97 & 9.07 & 0.49 & -2.24 & -2.57 & 164.95 & 19.67 & 167.79 & 1887 & 53.57 & 40.42 & 85.65 & 63.32 & -1135.88 & -848.39 & 0.01\\ \hline
14855 & -1814.25 & 94.00 & 148.98 & 11.33 & 0.49 & -1.14 & -1.15 & 163.44 & 19.76 & 173.71 & 1777 & 52.79 & 41.20 & 84.46 & 64.52 & -1124.67 & -863.29 & -0.01\\ \hline
EE & -1825.45 & 94.00 & 148.95 & –– & –– & –– & –– & 161.71 & 19.69 & 164.46 & 1984 & 53.68 & 40.31 & 85.68 & 63.26 & -1140.67 & -849.24 & -0.02\\ \hline
\end{tabular}
\caption{$^{243}$Pu Run 08 Asymmetric}
%\end{table}
%\begin{table}[h]
\centering
\begin{tabular}{|c|c|c|c|c|c|c|c|c|c|c|c|c|c|c|c|c|c|c|}
\hline
Flip & $E$ & $Z$ & $N$ & $E_\text{QP}$ & $n_k$ & $\langle j_z \rangle$ & $j_z$ & $Q_{20}$ & $Q_{30}$ & TKE & $t_\text{scission}$ & $Z_\text{HFF}$ & $Z_\text{LFF}$ & $N_\text{HFF}$ & $N_\text{LFF}$ & $E^\text{intr}_\text{HFF}$ & $E^\text{intr}_\text{LFF}$ & $\Delta E$ \\ \hline 
0 & -1826.45 & 94.00 & 149.16 & 0.84 & 0.42 & 2.46 & 2.92 & 179.72 & 24.30 & 174.08 & 1784 & 52.00 & 42.00 & 83.45 & 65.71 & -1114.70 & -885.86 & -0.03\\ \hline
2 & -1825.16 & 94.00 & 148.93 & 0.91 & 0.54 & 0.93 & 0.93 & 179.58 & 24.30 & 173.36 & 1693 & 51.62 & 42.38 & 83.07 & 65.85 & -1107.59 & -890.95 & -0.01\\ \hline
7202 & -1819.32 & 94.00 & 149.00 & 7.09 & 0.50 & -0.47 & -0.90 & 181.01 & 24.08 & 174.07 & 1571 & 51.82 & 42.18 & 82.95 & 66.05 & -1106.82 & -886.58 & -0.01\\ \hline
EE & -1826.43 & 94.00 & 149.00 & –– & –– & –– & –– & 179.60 & 24.29 & 172.06 & 1590 & 51.91 & 42.09 & 83.26 & 65.74 & -1111.93 & -886.59 & -0.02\\ \hline
\end{tabular}
\caption{$^{243}$Pu Run 09 Asymmetric}
%\end{table}
%\begin{table}[h]
\centering
\begin{tabular}{|c|c|c|c|c|c|c|c|c|c|c|c|c|c|c|c|c|c|c|}
\hline
Flip & $E$ & $Z$ & $N$ & $E_\text{QP}$ & $n_k$ & $\langle j_z \rangle$ & $j_z$ & $Q_{20}$ & $Q_{30}$ & TKE & $t_\text{scission}$ & $Z_\text{HFF}$ & $Z_\text{LFF}$ & $N_\text{HFF}$ & $N_\text{LFF}$ & $E^\text{intr}_\text{HFF}$ & $E^\text{intr}_\text{LFF}$ & $\Delta E$ \\ \hline 
0 & -1824.80 & 94.00 & 148.73 & 0.87 & 0.63 & 0.65 & 0.65 & 211.39 & 1.59 & 150.35 & 6160 & 51.19 & 42.81 & 82.19 & 66.55 & -1074.91 & -900.23 & 0.01\\ \hline
9006 & -1820.29 & 94.00 & 149.19 & 7.60 & 0.40 & -0.65 & -1.47 & 213.22 & 1.30 & 153.36 & 5384 & 50.35 & 43.65 & 81.06 & 68.13 & -1058.44 & -915.17 & 0.04\\ \hline
14428 & -1817.78 & 94.00 & 149.11 & 9.70 & 0.45 & 0.32 & 0.32 & 215.62 & 1.26 & 150.51 & 5326 & 50.79 & 43.21 & 82.03 & 67.08 & -1064.83 & -903.46 & 0.00\\ \hline
EE & -1826.97 & 94.00 & 149.00 & –– & –– & –– & –– & 211.65 & 1.59 & 151.62 & 4789 & 50.72 & 43.28 & 81.23 & 67.77 & -1065.66 & -912.92 & 0.01\\ \hline
\end{tabular}
\caption{$^{243}$Pu Run 00 Symmetric}
%\end{table}
\end{sidewaystable} 
\providecommand{\selectlanguage}[1]{}
\renewcommand{\selectlanguage}[1]{}

\clearpage 

\bibliography{local_fission}